\documentclass[lettersize,journal]{IEEEtran}
\usepackage{amsmath,amsfonts}
\usepackage{hyperref}
\usepackage{algpseudocode}
\usepackage{algorithm2e}
\SetArgSty{textup}
\usepackage{array}
\usepackage[caption=false,font=normalsize,labelfont=sf,textfont=sf]{subfig}
\usepackage{textcomp}
\usepackage{stfloats}
\usepackage{url}
\usepackage{verbatim}
\usepackage{graphicx}
\usepackage{cite}
\usepackage{mathtools}
\usepackage{float}\usepackage{xcolor}

\newcommand\hl[1]{%
  \bgroup
  \hskip0pt\color{black}%
  #1%
  \egroup
}
\DeclareMathOperator*{\argmin}{arg\,min}
\graphicspath{ {./Figures/} }
\begin{document}

\title{Hearing-Loss Compensation Using Deep Neural Networks: A  Framework and Results From a Listening Test}

\author{Peter Leer, Jesper Jensen, Laurel H. Carney ~\IEEEmembership{Member}, Zheng-Hua Tan, ~\IEEEmembership{Senior Member, IEEE,} Jan Østergaard, ~\IEEEmembership{Senior Member, IEEE,} Lars Bramsløw% <-this % stops a space
        
\thanks{This work is partly supported by Innovation Fund Denmark case number: 0153-00091B}% <-this % stops a space
\thanks{Audio demonstrations can be found at \url{https://p-leer.github.io/DNN-HLC_sounds/}.}
\thanks{P. Leer is with Eriksholm Research Centre (Part of Demant), 3070, Snekkersten, Denmark, and with the Department of Electronic Systems,
Aalborg University, 9220 Aalborg Øst, Denmark (e-mail: pelb@eriksholm.com).}%
\thanks{J. Jensen is with Eriksholm Research Centre (Part of Demant), 3070, Snekkersten, Denmark, and with the Department of Electronic Systems,
Aalborg University, 9220 Aalborg Øst, Denmark (e-mail: jesj@demant.com).}%
\thanks{L. H. Carney is with the Departments of Biomedical Engineering and Neuroscience, University of Rochester, 14642 Rochester, NY, USA (e-mail:  Laurel\_Carney@URMC.Rochester.edu)}
\thanks{J. Østergaard is with the Department of Electronic Systems,
Aalborg University, 9220 Aalborg Øst, Denmark (e-mail: jo@es.aau.dk).}%
\thanks{Z-H. Tan is with the Department of Electronic Systems,
Aalborg University, 9220 Aalborg Øst, Denmark Denmark, and also with the Pioneer Centre for
AI, 1350 Copenhagen, Denmark (e-mail: zt@es.aau.dk).}%
\thanks{L. Bramsløw is with Eriksholm Research Centre (part of Demant), 3070, Snekkersten, Denmark (e-mail: labw@eriksholm.com).}}%

\markboth{}%
{Shell \MakeLowercase{\textit{et al.}}:}

\IEEEpubid{}

\maketitle

\begin{abstract}
%\boldmath
This article investigates the use of deep neural networks (DNNs) for hearing-loss compensation. Hearing loss is a prevalent issue affecting millions of people worldwide, and conventional hearing aids have limitations in providing satisfactory compensation. DNNs have shown remarkable performance in various auditory tasks, including speech recognition, speaker identification, and music classification. In this study, we propose a DNN-based approach for hearing-loss compensation, which is trained on the outputs of hearing-impaired and normal-hearing DNN-based auditory models in response to speech signals. First, we introduce a framework for emulating  auditory models using DNNs, focusing on an auditory-nerve model in the auditory pathway. We propose a linearization of the DNN-based approach, which  we use to analyze the DNN-based hearing-loss compensation. Additionally we develop a simple approach to choose the acoustic center frequencies of the auditory model used for the compensation strategy. Finally, we evaluate, to our knowledge for the first time, the DNN-based hearing-loss compensation strategies using listening tests with hearing impaired listeners. The results demonstrate that the proposed approach results in feasible hearing-loss compensation strategies. Our proposed approach was shown to provide an increase in speech intelligibility versus an unprocessed baseline  and was found to outperform a conventional approach in terms of \hl{both intelligibility} and preference.
\end{abstract}

% === KEYWORDS ====================================================================
% =================================================================================

 \IEEEpubidadjcol 
 
\section{Introduction}
\label{sec:JP2_introduction}
Hearing loss is a debilitating condition, affecting almost 500 million people worldwide \cite{Rasiah2018AddressingLoss}. Hearing loss is usually managed by prescription of hearing aid (HA) devices that amplify acoustic sounds to make them audible to the afflicted patients. These devices usually consist of a system that can provide noise reduction (NR) to reduce the effect of background noise and competing speakers, followed by compressive amplification, that attempts to restore the loudness perception within the limited dynamic range of a HA user with impaired hearing. While significant effort has been invested in developing DNN-based noise reduction (NR) systems for hearing aids, such as those developed in the Clarity Challenge \cite{Akeroyd2020LaunchingProcessing,Akeroyd2023ResultsDevices,Cox2023OverviewAids}, there has been less focus on developing DNN-based hearing-loss compensation (HLC) strategies. In some cases, joint HLC and NR systems were trained on simplified auditory models, yet no dedicated HLC strategies have resulted from the Clarity Challenge.

\indent In recent years, auditory models have been developed that simulate the complex, non-linear behavior of neural representations of the auditory pathway when excited by acoustic stimuli. Although these models have been able to represent increasingly complex auditory phenomena, it is not clear how these representations can be translated into practical signal-processing strategies for hearing aids, as the neural representations are highly non-linear and extremely computationally expensive to generate.

\indent Recently, a methodology has been proposed, where DNNs are trained as emulators of auditory models, which are computationally expensive to evaluate using conventional solvers, leading to a decrease in inference time by several orders of magnitudes \cite{Nagathil2021ComputationallyProcessing,Nagathil2023WaveNet-basedModel,Baby2021AApplications,Drakopoulos2021ASynapses}. This methodology allows for the use of the DNN-based auditory model emulators (AMEs) for the purpose described in this work, namely developing HLC strategies using the outputs of the AMEs as biologically inspired optimization objectives. Ideally, this optimization restores the output of the hearing-impaired representation to that of normal hearing, hypothesizing that this strategy will enhance the hearing abilities of hearing-impaired individuals, as compared to conventional HLC strategies. Such an approach, if successful, could allow for development of not only perceptually relevant signal-processing strategies, but also the personalization of a HLC algorithm to an individual's particular dysfunction of the auditory pathway \cite{Biondi1978AuditoryViewpoint,
Bondy2004, Chen2005AMethod, Hengel2015SimulatingAids}.

In \cite{Drakopoulos2023a}, a DNN-based HLC strategy was proposed to compensate for mild hearing losses and synaptopathy using a biophysical model of the auditory nerve \cite{Verhulst2018}. The effectiveness of this method was evaluated using conventional objective metrics such as STOI \cite{Taal2011AnSpeech} and HASPI \cite{Kates2021The2}, though no subjective evaluations were conducted. A similar approach was taken in \cite{Drgas2024DynamicCompensation}, which utilized a simplified auditory model to compare several DNN-based HLC architectures using variations of STOI and HASPI, again without incorporating any subjective evaluation. However, it is crucial to recognize that conventional metrics, such as STOI, are not designed to account for hearing loss and their accuracy in such situations is questionable. Also, hearing-loss specific metrics, such as HASPI, have yet to be validated for DNN-based HLC strategies. 
In this work, we propose a framework for developing auditory-model driven HLC and joint HLC and NR strategies for mild to severe hearing losses. To do so, we build upon the DNN architecture and framework developed in \cite{LeerEtAl1}, which allows for emulation of auditory models spanning a large range of hearing losses, from mild to severe, and a corresponding large range of input signals, including significantly amplified signals. 
Unlike existing work, which evaluates the effectiveness of the proposed methods using objective metrics, we instead evaluate our approach using subjective listening tests on a cohort of hearing-impaired listeners. Considerations related to causality, latency, and memory requirements are intentionally excluded; if an improvement in intelligibility over conventional approaches can not be observed in a general unconstrained DNN, it is unlikely that more constrained, resource-efficient DNNs will yield improved results.

The article is structured as follows: We give a short introduction to conventional hearing loss compensation schemes in Sec. \ref{sec:JP2_conv_HLC}. 
In Sec. \ref{sec:JP2_AME} we review a framework to emulate an auditory-nerve model using DNNs. A linearized model of an auditory-model based hearing-loss compensation is introduced in Sec. \ref{sec:JP2_aud_based_HLC}, which allows us to analyze the linear effects of the DNN-based compensation strategy and choose hyperparameters for finding the final DNN-based HLC strategy. In Sec. \ref{sec:JP2_comp_networks} we introduce the DNN architectures used for HLC and NR and how we train these networks. The listening experiments are presented in Sec.\ref{sec:JP2_experiments}, and evaluated in \ref{sec:JP2_results} where we compare a conventional HLC and NR to our proposed auditory-model based HLCs and joint HLC and NR strategies, using listening experiments conducted on hearing impaired subjects. In Sec. \ref{sec:JP2_Discussion} we discuss the results and their implications, possible limitations and perspectives for future work. Finally, in Sec. \ref{sec:JP2_conclusion} we provide a short overview of our findings, concluding our work.
\section{Conventional hearing-loss compensation strategies}
\label{sec:JP2_conv_HLC}
Conventional hearing-loss compensation strategies can usually be grouped into either loudness-normalization rationales or loudness-equalization rationales. The primary goal of a loudness-normalization rationale is to restore the loudness perception in each frequency band. A "soft", "medium" or "loud" narrow band sound for a normal hearing individual should be perceived equivalently as a "soft", "medium" or "loud" narrow band sound for the HA user wearing a hearing aid. Secondarily, the relative loudness between frequency bands should not change. The goal of a loudness-equalization rationale is, on the other hand, that the HA user should perceive all frequency bands as equally loud, which is argued to increase speech intelligibility \cite{K.V.Lindley2002}. One such  loudness-equalization rationale is the NAL-Revised (NAL-R) \cite{Byrne1986TheAid}, that provides a linear time-invariant gain, parameterized by the pure-tone audiogram of the HA user.
However, because shifts in hearing thresholds often are not accompanied by similar shifts in thresholds of discomfort for loud sounds, hearing impairment results in a reduced dynamic range. 
To address this, modern hearing aids use wide dynamic range compression to compensate \cite{LindleyIV1997FittingAids}. Multi-band compression is commonly employed to provide precise control over distinct frequency bands, minimizing the impact of transients on lower-frequency speech cues and matching varying input dynamic ranges across frequencies \cite{Schneider1997MultichannelAid}\cite{Kates2005MultichannelWarping}.

  \section{Auditory-model emulation}
\label{sec:JP2_AME}
In this section we review the auditory model used for in, the Zilany model \cite{Zilany2014}, and how we emulate it using deep neural networks. Additionally, we review the Verhulst auditory model\cite{Verhulst2018}, which we used for our linear-model analysis in Sec. \ref{sec:JP2_aud_based_HLC}.

First, we define the generic notation that will be used when describing auditory models in this work. Define an acoustic signal space $X \subset \mathbb{R}^T$, and an inner representation space $I \subset \mathbb{R}^{K\times T}$, where $K$  denotes the number of auditory-model frequency channels and $T$ denotes the number of time-samples. The auditory model is defined as $f^\theta : X \rightarrow I$, where $\theta$ is a vector containing the free parameters of the auditory model. Additionally, when we set $\theta$ to "NH", it represents normal hearing parameters; when set to "HI", it represents hearing-impaired parameters in general. We denote the output of the $k$-th channel as $f_k(x)$ with $x\in X$.

\subsection{Overview of the Zilany model and implementation choices}
\begin{figure}
  \centering
      \includegraphics[scale=0.35]{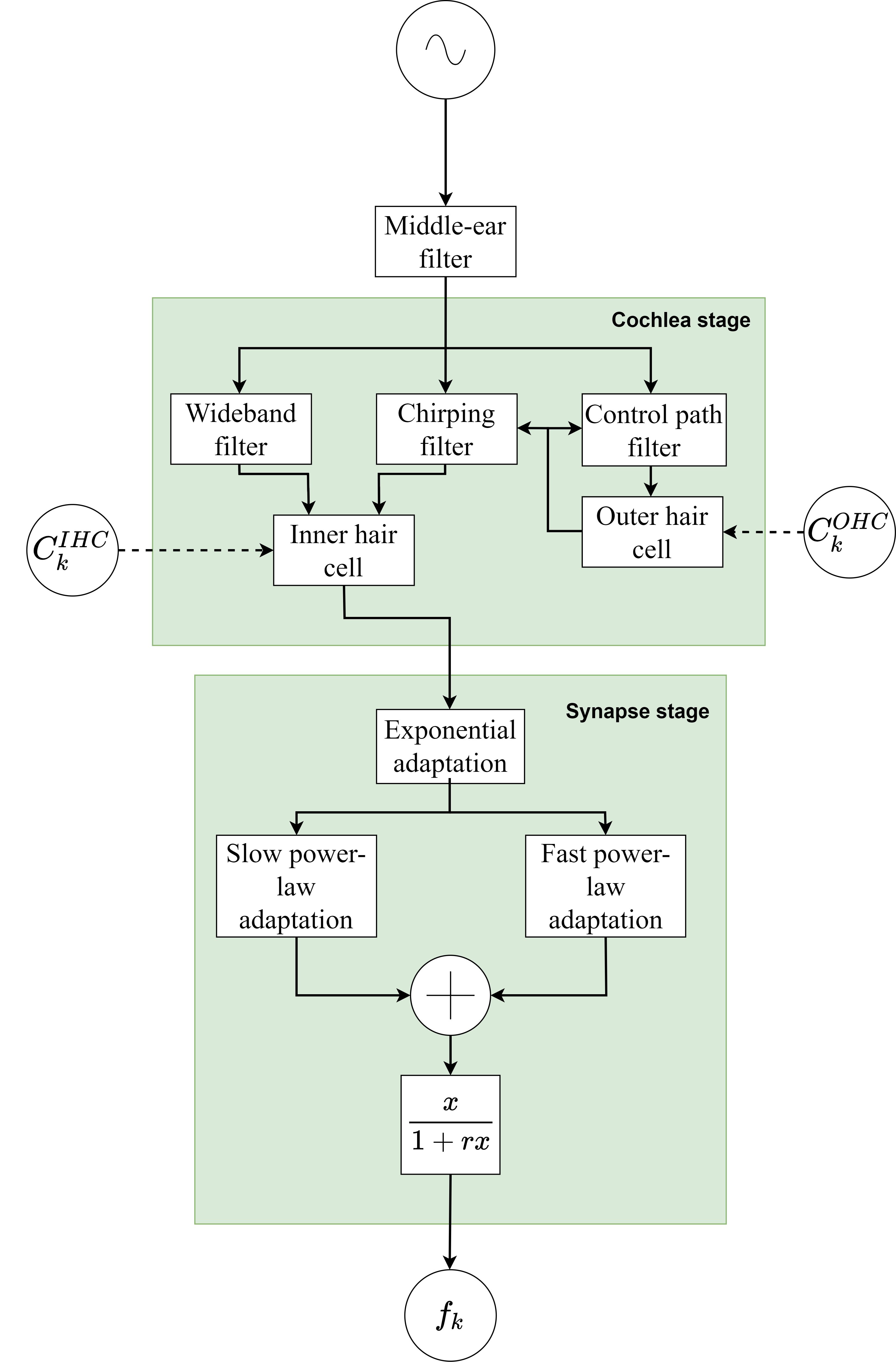}
  \caption{Overview of the Zilany model }
  \label{fig:zilanyoverview}
\end{figure}

The Zilany model is an auditory model that simulates the auditory-nerve response to an acoustic signal \cite{Zilany2014}. The model comprises multiple stages of the auditory pathway, including a middle-ear stage, a combined cochlea stage that mimics the behavior of outer hair cells (OHCs) and inner hair cells (IHCs), and an IHC-to-auditory-nerve synapse stage, used to model the expected firing rate of the auditory nerve. An implementation of the model can be found in \cite{AuditoryCenter}, and an overview of the model can be seen in Fig. \ref{fig:zilanyoverview}.
 The Zilany model can be partitioned into the following stages:

\subsubsection{Middle-ear stage}
The middle-ear stage is a linear circuit model, implemented as a set of cascaded filters, that simulates the forward transmission from the ear canal to the oval window.

\subsubsection{Cochlea stage}
This stage consists of two branches: A chirping filter, where the gain and bandwidth of the filters are controlled by the input sound pressure level (SPL) through the control signal path, simulating the level-dependent properties of the cochlea, and a linear, static filter that dominates at high input SPLs. The outputs of both filters are fed through their individual transduction functions, such that they combine to generate the IHC potential. Each auditory-model channels includes two free parameters, denoted as $C^{OHC}_k \in  [0,1]$ and $C^{IHC}_k \in  [0,1]$, representing the state of the OHCs and IHCs. A value of 0 denotes complete dysfunction, while 1 represents normal hearing. 
The loss of OHCs introduces elevated thresholds, broadened tuning and a reduction of compression and suppression. The loss of IHCs contributes to elevated thresholds in the model.

\subsubsection{Synapse stage}
The IHC-to-auditory-nerve synapse stage is a power-law adaptation model, consisting of an exponential adaptive stage that drives two parallel power-law adaption paths: A slow path and a fast path \cite{Zilany2009ADynamics}. In the original model, fractional noise is used to model the time-varying distribution of the spontaneous rate. However, in this work, we are only interested in the deterministic response to stimuli, particularly speech,  and thus this noise process was omitted. Finally, the output of the synapse, is processed by a rational function, that converts the synapse output to an expected value of the time-varying rate function of the auditory nerve. For this work, we simulate auditory-nerve fibers with low, medium and high spontaneous rates, corresponding to different synaptic characteristics. We use an equal number of low-, mid-, and high-rate fibers, concatenated together, assuming no fiber loss.
For each of the fibers, we computed a reference response with no input, resulting in the spontaneous activity that is not driven by the fractional Gaussian noise. The reference response was subtracted from the response to any stimulus, resulting in a firing rate that was zero when there was no input.

In order to model hearing losses in the Zilany model, a function, "fitaudiogram2", is provided with the auditory model code \cite{AuditoryCenter}. This function takes an audiogram as input and produces $C^{OHC}_k$ and $C^{IHC}_k$for each auditory-model channel with corresponding center frequency (CF). Because there is no way to accurately measure the relative contribution of OHC and IHC loss, we distribute 2/3 of the hearing loss to OHCs and 1/3 to IHCs for all audiograms, which is the default setting in the "fitaudiogram2" function. This distribution corresponds to the maximal threshold shift attributable to OHC damage \cite{Bruce2003AnResponses}. OHC loss often dominates over IHC loss in cases of hearing impairment, a pattern observed in various studies \cite{Wu2019PrimaryEar, Wu2021PrimaryScores}. Additionally, we smooth $C^{OHC}_k$ and  $C^{IHC}_k$ across CFs using a 3-point moving average. 
\subsection{The Verhulst model}
The Verhulst model is an auditory model that consists of a transmission-line model of the basilar membrane, an IHC stage, an auditory-nerve stage, and a midbrain stage \cite{Verhulst2018}. We use this model to perform a linear analysis of a previously derived DNN-based HLC \cite{Drakopoulos2023ACompensation}. Hearing loss can be emulated by parameterizing the non-linearities of the outer hair cells in the model, such that the amplitude response to a single tone, at a given CF, is reduced by the specified loss. The model only captures hearing loss attributed to OHCs, which is maximally 35 dB at every CF\cite{GitHubHearingTechnology/CoNNear_cochlea}, and synaptopathy (permanent loss of synapses in the auditory nerve), which can be modeled by reducing the number  of nerve fibers in the auditory-nerve stage. Since there is no established way to parameterize larger, clinically relevant hearing losses in the Verhulst auditory model, we generate the final HLC and NR strategies on the Zilany model, cf. Sec. \ref{sec:JP2_comp_networks}. 
\subsection{Auditory model emulators}
In order to significantly decrease the computational time required for the auditory model to process an input signal (the forward pass), and the computation of the partial derivative of the model with respect to the input signal (the backward pass), we emulate the auditory model using a DNN, which we call an auditory-model emulator (AME). 

In general, training an AME presents several challenges \cite{LeerEtAl1}, including:
\begin{enumerate}
    \item Initial pilot-experiments found that the auditory models were sensitive to the choice of hyperparameters. Thus, finding suitable AME architectures and hyperparameters is non-trivial \cite{Baby2021AApplications}.
    \item  The DNN must be able to emulate the auditory model across all auditory-model channels, different types of signals and relevant sound pressure levels, which can not be achieved using conventional loss functions \cite{LeerEtAl1}.
    \item Conventional training methods of deep neural networks induce a bias towards producing a low-frequency representation of the target function, meaning that certain output channels of the auditory-model emulator will converge faster to the ground-truth auditory model, even if all the frequency components are normalized\cite{Rahaman2018OnNetworks}. This behavior was verified in our pilot experiments, and in some scenarios the high-frequency channels would not converge.
\end{enumerate}
Thus, these three issues had to be addressed during training.  We addressed these issues individually in the following way:
\subsubsection{Finding a suitable architecture}
Finding a suitable architecture for an AME is largely an empirical process. There are some constraints, such as having a large enough receptive field to properly model the time constants of the original auditory model. Feed-forward convolutional models are good candidates, since they impose a translational equivariance prior on the DNN \cite{Cohen2016GroupNetworks}, a property that aligns with the input-output relations of auditory models. This helps to reduce the number of parameters needed, and allows for fast backpropagation. In particular, an auto-encoder structure allows for efficient representations of the long time constants of the auditory model, which led us to choose the Wave-U-Net structure, a convolutional auto-encoder \cite{Stoller2018}. The hyperparameters are shown in Table \ref{tab:WUN}. We found that different activation functions could lead to significantly varying performance, also shown in previous work \cite{Baby2021AApplications,Drakopoulos2021ASynapses}. Based on experimental work, we chose the hyperbolic tangent (Tanh) activation in the encoder (all the downsampling blocks) and the PReLU activation \cite{He2015} in the decoder (all the upsampling blocks and the embedding block). With the hyperparameters shown in Table \ref{tab:WUN}, we achieved a receptive field corresponding to roughly 0.25 seconds.
\subsubsection{Ensuring equal performance across channels and inputs}
To achieve good performance across auditory-model channels and inputs at different SPLs, we train the network using the frequency and level-dependent mean-absolute error (FMAE) \cite{LeerEtAl1}, which normalizes the error across input levels and auditory model channels. \hl{ First, define an estimate of $f$ as $\hat{f}$, and the output of the AME as $\bar{f}$. Denote $\textbf{x}_l \in X$ as $\textbf{x}$ at the $l$-th SPL of the training dataset and define $\alpha_{k,l}$ and $\beta_k$ as parameters related to the distribution of energy of the output of the auditory models \cite{LeerEtAl1}.  Then, FMAE is defined as:      
\begin{equation}
\resizebox{0.45\textwidth}{!}{$
    \mathrm{FMAE}(f(\textbf{x}_l),\bar{f}(\textbf{x}_l); \beta_k,\alpha_{k,l}) = \dfrac{1}{T K
    } \sum\limits_{k=1}^K||\beta_k f_{k}(\textbf{x}_l) -\bar{f}_{k}(\textbf{x}_l)||_1 \alpha_{k,l}  \, , 
    \label{eq:JP2_FMAE}
    $}
\end{equation}
where $T$ is the size of the vector representing the output of each auditory-model channel, and:
\begin{equation}
    \hat{f}_k(\textbf{x}) =\dfrac{\bar{f}_k(\textbf{x})}{\beta_k} \, .
\end{equation}
}
\subsubsection{Addressing the frequency bias of the AME} 
To address the issue of frequency bias in the training of the AME, we started with small batch sizes for faster convergence, and increased the batch size if the validation error did not increase for 5 epochs. The batch size was increased until the capacity of the GPU memory was reached (24 GB). We found, empirically, that the initial small batch size enabled training of the model within a reasonable time-frame, while increasing the batch size as the model was closer to convergence helped decrease the frequency bias.

\iffalse
\begin{figure}
  \centering
      \includegraphics[scale=0.45]{Wave_U_Netdrawio.png}
  \caption{Overview of the Wave-U-Net \cite{Stoller2018} architecture used for emulating the Zilany model. Ptw. denotes pointwise.}
  \label{fig:WUN}
\end{figure}
\fi

\begin{table}[H]
\centering
      \caption{Hyperparameters of the AME. N denotes the number of Upsampling/Downsampling blocks.}
      \label{tab:WUN}
\begin{tabular}{|l|l|}
\hline
N     & 8   \\ \hline
Kernel size                 & 21     \\ \hline
Channels & 128 \\ \hline
Decimation factor & 2 \\ \hline
Encoder activation & Tanh  \\ \hline
Decoder activation & PReLU \\ \hline
Bias & None\\ \hline
\end{tabular}

\end{table}
\subsection{Training of AMEs}
{An AME was trained for a particular hearing loss, using identical hyperparameters and speech dataset: the LibriTTS dataset\cite{Zen2019LibriTTS:Text-to-Speech}. The training set consisted of 4000 utterances and the validation set consisted of 500 utterances, chosen as a trade-off between computational resources, storage and computation time. $K=128$ CFs were used, with the spacing found by Algorithm \ref{algo:CF_spacing}, explained in Sec. \ref{sec:JP2_spacing}. \hl{The models were evaluated using the signal-to-error ratio (SER), defined as the energy ratio of the ground-truth output to the estimation error. Each auditory model channel achieved at least 10 dB SER \cite{LeerEtAl1} on a smaller test set of 20 sentences from LibriTTS.}

The auditory models were trained with inputs normalized to SPLs between 55 and 95 dB SPL, using signal-to-noise ratios (SNRs) covering noisy speech to essentially clean speech, specifically $\{-3,0,3, 6, 9, 12, \text{clean}\}$ dB SNR, where clean denotes the SNR of the original speech dataset, e.g. data without any additive noise added. In order to increase the robustness to relevant noise types and various signal modifications, we also applied the following augmentations to the input signals used to train the AME: Addition of babble noise and speech-shaped Gaussian noise at a large range of SNRs and band-pass filtering with either band-pass or high-pass filtering of the noisy speech, using random widths and center frequencies. Additionally, in order to expose the AME to a realistic hearing-loss compensation strategy, we processed some of the inputs using NAL-R \cite{Byrne1986TheAid}.}

 \section{Analysis of auditory-model-based hearing-loss compensation strategies}
\label{sec:JP2_aud_based_HLC}
In this section we review how hearing-loss compensation (HLC) strategies based on auditory models can be derived. Based on this, we derive linear HLC strategies, which we use as a basis of comparison with both a previously proposed DNN-based HLC strategy \cite{Drakopoulos2023ACompensation}  and our proposed DNN-based HLC strategy. We show that the previously proposed strategy \cite{Drakopoulos2023ACompensation} suffers from an undesirable ripple-like structure in the long-term frequency response. Interestingly, we find that this behavior can be predicted by the linear HLC strategies. We show that the ripple structure stems from the auditory-model channel spacing used to derive the DNN-based HLCs and provide an algorithm that reduces the ripple.

\subsection{Deriving auditory-model based hearing-loss compensation strategies}
\begin{figure*}
  \centering
      \includegraphics[scale=0.4]{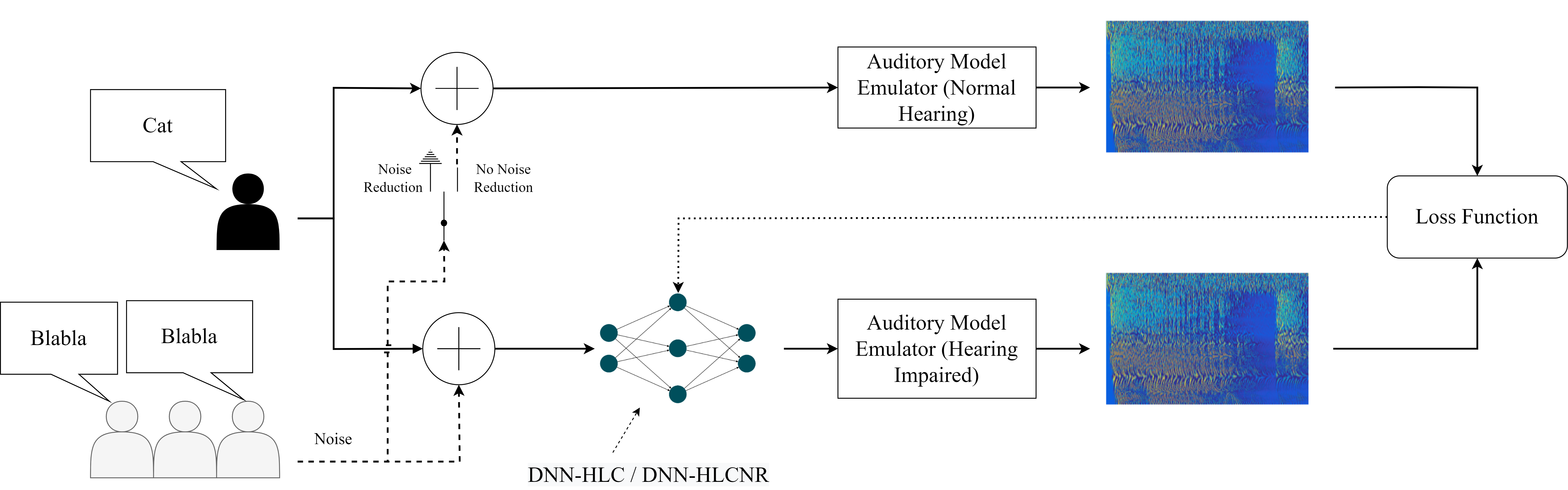}
  \caption{Training setup for a deep-neural-network based hearing-loss compensation (DNN-HLC / DNN-HLCNR) is trained.  In each step, speech material passes through a normal-hearing auditory-model emulator (AME). Noise is added both to the upper branch (the normal-hearing AME) and the lower branch when training the DNN-HLC, to ensure that the network performs well in noisy speech. Noise is only added to the the lower branch when training the DNN-HLCNR. The noisy speech material is simultaneously fed through the DNN-HLC(NR), and the output of the DNN-HLC(NR) is fed through a hearing impaired auditory model emulator. The outputs of the two AMEs are compared through a loss function, which is minimized by iteratively updating the weights of the DNN-HLC(NR), through backpropagation.}
  \label{fig:framework}
\end{figure*}

In general, auditory-model based hearing-loss compensation strategies are derived by finding a transformation that minimizes the dissimilarity between the output of a normal-hearing auditory model, whose input is an uncompensated signal, and the output of a hearing-impaired auditory model, whose input is a hearing-loss compensated signal, cf. Fig.  \ref{fig:framework}. 
We call this transformation an optimal compensation strategy with respect to a given dissimilarity measure.
\hl{
Define a compensation strategy as a function $c: X \rightarrow X$.
To find an optimal compensation strategy, find the function $c(\cdot)$ that minimizes:
\begin{equation}
    L(f^{\textrm{NH}}(\textbf{x}),f^{\textrm{HI}}(c(\textbf{x}))) \, ,
\end{equation}
where $L: I \times I \rightarrow \mathbb{R}^+ $ is some measure of dissimilarity ---  a loss function in optimization terminology --- and $\textbf{x}$ is an acoustic signal, $\textbf{x} \in X$, sampled from a distribution with finite second order moments.

\subsection{Optimal linear 
compensation strategies}

\label{Sec:opt_comp_strategies}
In order to find optimal, linear compensation strategies we can consider the conventional mean-squared error (MSE) as a loss function, $L(\cdot)$, because this loss function admits simple, closed-form solutions for linear and time-invariant systems.

Define the impulse response (or inverse Fourier transforms of the excitation patterns) of the $k$-th channel of the normal-hearing auditory model as  $\textbf{h}_k \in \mathbb{R}^{T \times 1}$. Responses for consecutive auditory-model channels can be assembled into the matrix:
\begin{equation}
    \textbf{N} =  \begin{bmatrix}
        \textbf{h}_1 & \textbf{h}_2 & \hdots & \textbf{h}_K
    \end{bmatrix}^T \in \mathbb{R}^{K \times T} \, .
\end{equation}
Similarly, construct the matrix $\textbf{D} \in \mathbb{R}^{K \times T} $ using $\textbf{h}_k$ derived from the hearing-impaired auditory model. We consider a linear, time-invariant compensation strategy, which can be expressed as a convolution between the input acoustic signal, $x \in \mathbb{R}^L$, and a compensation filter, $\textbf{c} \in \mathbb{R}^M$:
\begin{equation} 
c(\textbf{x}) = \textbf{c}  * \textbf{x} \, .
\end{equation}
To avoid time aliasing when performing convolutions and transformations to the frequency domain, we zero-pad $\textbf{c}$, \textbf{x} and each $\textbf{h}_k$ to a common length $N$, where $N\geq L+T+M-2$. Then, the convolution between $\textbf{x}$ and $\textbf{c}$ can be realized by applying the linear operator that transforms a vector, $ \textbf{x}$, into its Toeplitz matrix:
\begin{equation}
\mathbb{T}(\mathbf{x}) = \begin{bmatrix}
x_1      & 0        & \cdots & 0        & 0       \\
x_2      & x_1      &  \ddots & 0 & 0 \\ 
x_3      & x_2   
   & x_1      & \ddots & 0     \\
\vdots & \ddots & \ddots & \ddots & 0 \\ 
x_N      & \cdots   & \cdots & x_2      & x_1 
\end{bmatrix} \in \mathbb{R}^{N \times N} \, ,
\end{equation}
such that:
\begin{equation}
   \textbf{c}  * \textbf{x} = \mathbb{T}(\mathbf{x})  \textbf{c} \, . 
\end{equation}
}
Define the output of the linear, normal-hearing auditory model as the linear convolution between each row in \textbf{N} and the input signal, $\textbf{x}$:

\begin{align}    
f_{lin}^{NH}(\textbf{x}) &= \textbf{N} \mathbb{T}(\textbf{x})^T \in \mathbb{R}^{K \times N} \, .
\end{align}
Similarly, the output of the linear, hearing-impaired auditory model is given by:
\begin{align}    
f_{lin}^{HI}(c(\textbf{x})) &= \textbf{D} \mathbb{T}(c(\textbf{x}))^T \in \mathbb{R}^{K \times N} \, .
\end{align}

We find an MSE-optimal, linear compensation strategy, $\textbf{c}^*$, by minimizing the mean squared difference of the output of the two auditory models:
\begin{equation}
    \textbf{c}^* = \displaystyle{\argmin_{\textbf{c} \in \mathbb{R}^N}} \, \mathbb{E}_\textbf{x}\left[||\textbf{N} \mathbb{T}(\textbf{x})-\textbf{D} \mathbb{T}(\mathbb{T}(\textbf{x})\textbf{c}  )^T ||_F^2\right] \, .
    \label{eq:JP2_toep_MSE}
\end{equation}

\noindent To solve the problem efficiently and to ease interpretation, we transform the minimization problem to the discrete Fourier domain. Let $\textbf{W} \in \mathbb{C}^{N \times N} $ denote the discrete Fourier transform matrix, such that the discrete Fourier transform along each row of $\textbf{N}$ is given by $\textbf{N} \textbf{W} $, and the Fourier transform of a vector, $\textbf{x}$, is given by $\textbf{W}\textbf{x}$.
Define $\tilde{\textbf{N}} = \textbf{N}\textbf{W} $, $\tilde{\textbf{D}} = \textbf{D}\textbf{W}$, $\tilde{\textbf{x}} = \textbf{W}\textbf{x}$ and $\tilde{\textbf{c}} = \textbf{W}\textbf{c}$.  Then, using Parseval's theorem \cite{Oppenheim1998DiscreteEdition}, one can show that Eq. (\ref{eq:JP2_toep_MSE})
 is equivalent to:

\begin{equation}
    \tilde{\textbf{c}}^*= \displaystyle{\argmin_{\tilde{\textbf{c}} \in \mathbb{C}^N}} \,\mathbb{E} _{\tilde{\textbf{x}}}\left[
    ||\tilde{\textbf{N}} \textrm{diag}(\tilde{\textbf{x}})-\tilde{\textbf{D}}\textrm{diag}(\tilde{\textbf{x}})\textrm{diag}(\tilde{\textbf{c}})||_F^2 \right]\, ,
    \label{eq:JP2_frequency_MSE} 
\end{equation}

\noindent where diag$(\textbf{x})$ denotes the diagonal matrix with the vector, $\textbf{x}$, in the main diagonal.
Solving for $\tilde{\textbf{c}}$ leads to:
\begin{equation}
    \tilde{\textbf{c}}^* = (\tilde{\textbf{D}}^H \tilde{\textbf{D}} \odot \textbf{I})^{-1}  (\tilde{\textbf{D}}^H \tilde{\textbf{N}} \odot \textbf{I}) \textbf{1} \, ,
    \label{eq:JP2_optc}
\end{equation}
or, alternatively,
\begin{equation}
    \tilde{c}^*_i=\dfrac{(\tilde{\textbf{D}}^H \tilde{\textbf{N}})_{ii}}{(\tilde{\textbf{D}}^H \tilde{\textbf{D}})_{ii}}\, ,
\end{equation}
    with $(\cdot)_{ii}$ denoting the element at the i-th row and i-th column of the matrix, $(\cdot)^H$ the Hermitian transpose, $\odot$ the Hadamard product, $\textbf{I}$ the identity matrix  and $\textbf{1}$ the vector containing all ones. Eq. (\ref{eq:JP2_frequency_MSE}) implies that hearing loss can only be completely restored if $\tilde{\textbf{N}}= \tilde{\textbf{D}} \textrm{diag}(\tilde{\textbf{c}})$. This requires that all pairwise columns of $\tilde{\textbf{N}}$ and $\tilde{\textbf{D}}$ are proportional, that is all auditory filters in the normal and hearing-impaired model differ by the same scaling factor at each frequency bin. A particular situation, where this is trivially satisfied is for auditory-model filters with non-overlapping support. However, such a model would be highly non-realistic. Generally, the requirement, $\tilde{\textbf{N}}= \tilde{\textbf{D}} \textrm{diag}(\tilde{\textbf{c}})$, results in unrealistic auditory models and hearing losses, implying that realistic hearing losses can not be completely restored for the linear and time-invariant auditory model in a squared-error sense at the peripheral stage of the auditory pathway.
\
\subsection{Comparison of linear HLCs and DNN-based HLCs}
For any compensation strategy (linear or non-linear), we define the long-term linear frequency gain as:
\begin{equation}
\textbf{g} = \mathbb{E}_\textbf{x}[|\textbf{W}c(\textbf{x})|]\oslash \mathbb{E}_\textbf{x}[|\textbf{W}\textbf{x}|]\, ,
\end{equation} where $\oslash$ denotes component-wise division (for our experiments, we used Welch's method \cite{Welch1967} as an estimator of the expectation). In our initial experiments, where we trained DNNs to perform the compensation --- cf. Sec. \ref{sec:JP2_comp_networks} on details about the training procedure ---, we found a ripple structure in the resulting long-term gain, cf. the blue, dashed curve in Fig. \ref{fig:ripple_zilany}. We conjecture that the observed ripple is caused by certain frequencies dominating the loss function, resulting in increased gain at those frequencies: When auditory filters are sharply tuned relative to the hearing loss model and adjacent filters decay rapidly, local maxima occur close to each CF in Eq. (\ref{eq:JP2_optc}). This results in a comb-like ripple pattern in the gain function that depends on the spacing of the CFs of the auditory-model filters. Depending on the severity of the ripple, it can potentially cause audible artifacts.  We examine our conjecture through two scenarios: 1) Comparing the linear, predicted gain with a DNN-based compensation strategy based on the Zilany auditory model and 2) comparing the linear, predicted gain with a pre-trained DNN based on the Verhulst auditory model, using a different, composite loss function \cite{Drakopoulos2023ACompensation}. In particular, we focus on higher frequencies where the quality factors of the auditory model filters tend to increase, leading to increased ripple.
\subsubsection{Comparison of the linear model with a DNN-based HLC using the Zilany auditory model}
\begin{figure}[H]
  \centering
      \includegraphics[scale=0.5]{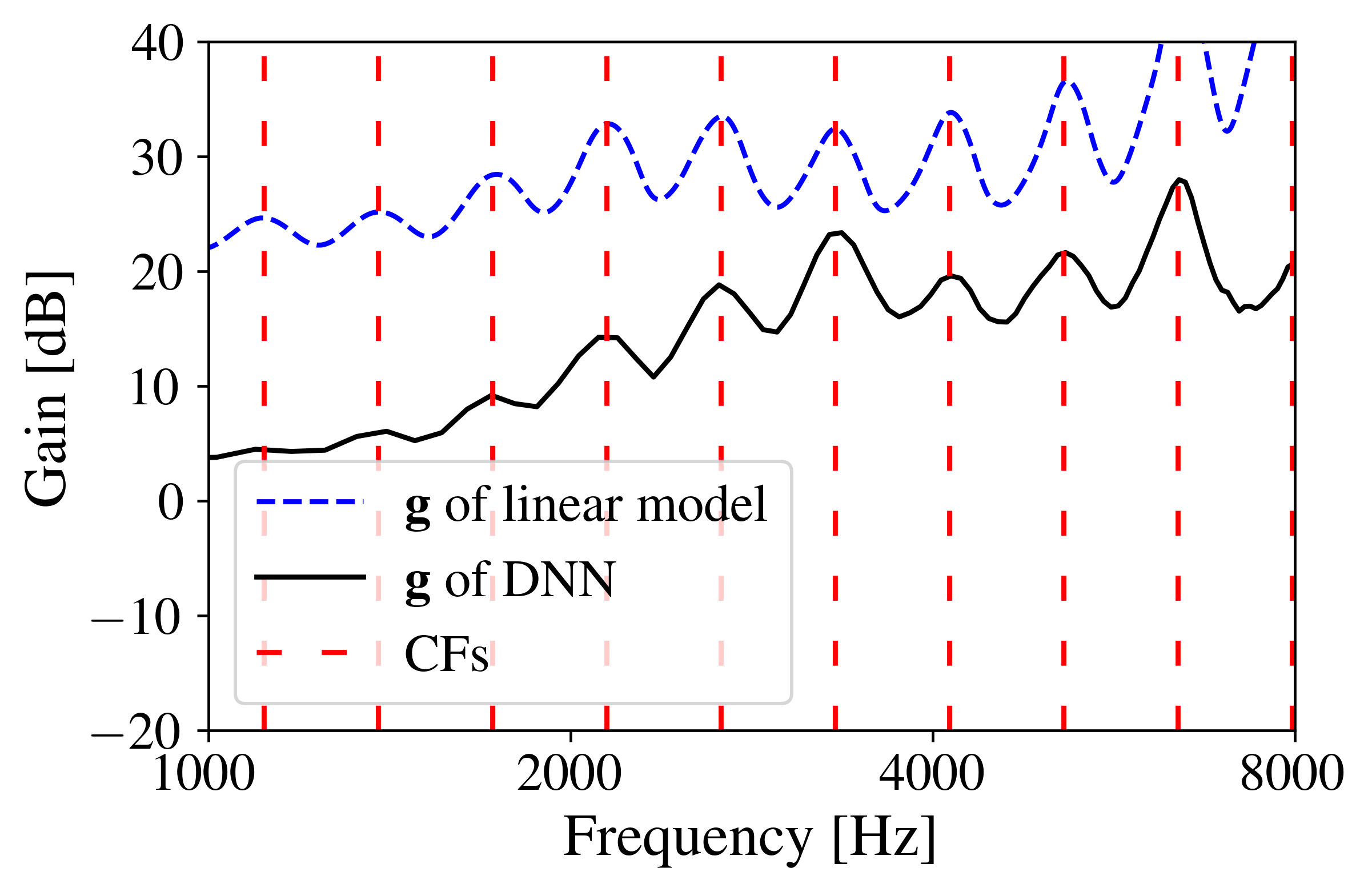}
  \caption{Long-term average gain of a DNN-based hearing loss compensation strategy, optimized at the auditory nerve level of the Zilany model and a linear, optimal gain, based on a 40 dB flat hearing loss. 21 log-spaced CFs were used to generate the figure}.
  \label{fig:ripple_comparison}
\end{figure}

In Fig. \ref{fig:ripple_comparison} we compare the gain of the linear HLC, based on Eq. (\ref{eq:JP2_frequency_MSE}), with the long-term frequency gain of a DNN-based HLC strategy, using a flat hearing loss of 40 dB. We see that the linear gain is able, to a large extent, to predict the ripple structure, i.e., it predicts the locations of the peaks, and to some degree the relative magnitude of the ripple. 
\subsubsection{Comparison of the linear model with a DNN-based HLC using the Verhulst auditory model}
\begin{figure}[H]
  \centering
      \includegraphics[scale=0.5]{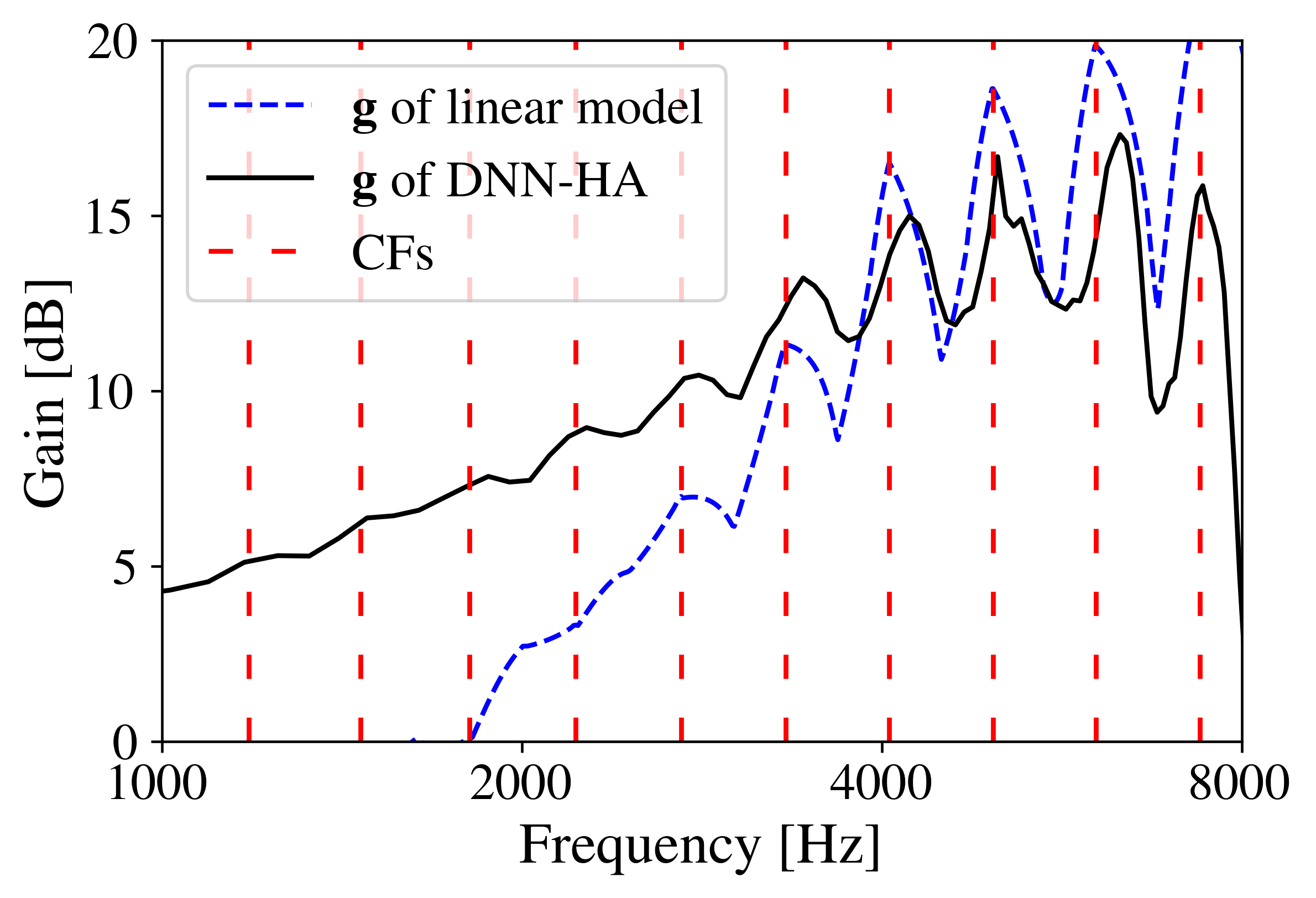}
  \caption{Long-term average gain of the DNN-HA, optimized at the auditory nerve level of the Verhulst model using a composite loss function and a linear, optimal gain, based on a 35 dB flat hearing loss.}
  \label{fig:fotis}
\end{figure}

In Fig. \ref{fig:fotis} we compare the gain of the linear HLC, based on Eq. (\ref{eq:JP2_frequency_MSE}), with the long-term frequency gain of a pre-trained DNN-based hearing-loss compensation strategy (DNN-HA) from \cite{Drakopoulos2023AAll}\footnote{The pre-trained DNN-HA can be found at \href{https://github.com/fotisdr/DNN-HA}{https://github.com/fotisdr/DNN-HA}}. DNN-HA is optimized using the Verhulst auditory model with a 35 dB flat hearing loss. The DNN was derived using 21 CFs and a composite loss function based on absolute differences of individual auditory nerve responses, population responses, both in the time-domain and after transforming the responses to their magnitude spectral representation. Additional terms were added to enhance temporal contrast while restricting processing frequencies to below 8 kHz, see \cite{Drakopoulos2023a} for details. Even though the composite loss function differs from the MSE-based optimization objective used to derive the linear gain, we find that the optimal, linear gain can, to some extent, predict the location and the occurrence of the ripple-structure of the signals processed by the DNN-HA.

\subsection{Choosing appropriate CFs}
\label{sec:JP2_spacing}
A spacing of auditory-model channels that reduces the ripple will depend on the particular auditory model and the amount of acceptable ripple. Thus, to construct a set of center frequencies  that results in reduced ripple, we propose an algorithm that recursively builds a list of center frequencies, by ensuring that the amplitude response at the next CF only decays a certain amount, $\Delta$, see Algorithm \ref{algo:CF_spacing}. By using the proposed algorithm, auditory model filters with higher CFs, which decay faster, are placed in closer to proximity to another as compared to e.g., log spacing.

\begin{figure}[H]
  \centering
      \includegraphics[scale=0.5]{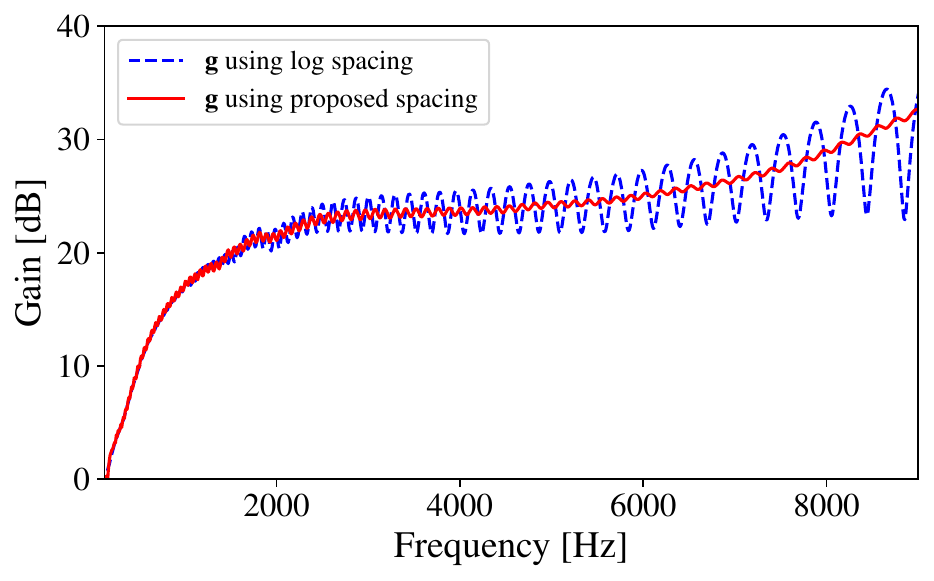}
  \caption{Comparison of optimal-compensation insertion gains for the output of the chirping filter of the Zilany auditory model, using K=96 CFs for log spacing and the proposed spacing
  and a moderate hearing loss, N3 \cite{Bisgaard2010}. The proposed spacing is generated by using Algorithm \ref{algo:CF_spacing}.}
  \label{fig:ripple_zilany}
\end{figure}

\begin{figure}[H]
  \centering
      \includegraphics[scale=0.5]{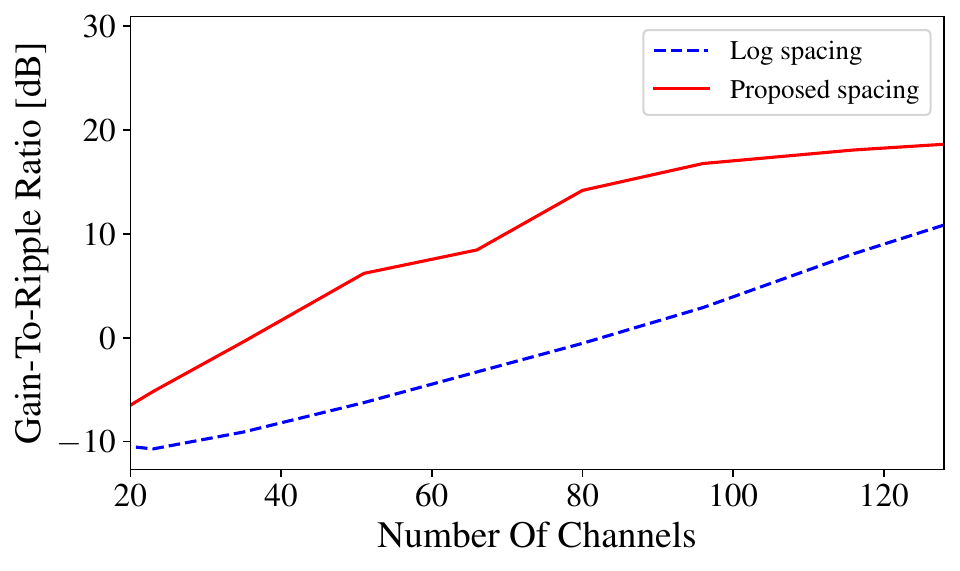}
  \caption{Gain-to-ripple ratio calculated using the output of the chirping filter in  the Zilany auditory model, using a N3 audiogram \cite{Bisgaard2010}, a moderate hearing loss. The CFs are generated with $\textrm{CF}_{min} = 125 \textrm{ Hz}$ and $\textrm{CF}_{max} = 9900 \textrm{ Hz}$. The proposed spacing is generated by Algorithm \ref{algo:CF_spacing}.}
  \label{fig:spacing}
\end{figure}

\begin{algorithm}
\SetKwInOut{Input}{Input}
\SetKwInOut{Output}{Output}
\Input{Minimum center frequency $\mathrm{CF}_{\min}$; Maximum center frequency $\mathrm{CF}_{\max}$; Threshold $\Delta$}
\Output{A list of center frequencies $\mathrm{CFs}$ spanning from $\mathrm{CF}_{\min}$ to $\mathrm{CF}_{\max}$}
\BlankLine
Initialize $v \leftarrow \mathrm{CF}_{\min}$\;
Initialize $\mathrm{CFs} \leftarrow$ empty list\;
\While{$v < \mathrm{CF}_{\max}$}{
    \begin{enumerate}
        \item Compute the frequency response $\mathbf{z}$ of the filter $f^{NH}$ with center frequency $v$:
        \[ \mathbf{z} = \mathrm{frequencyResponse}(f^{NH}, \mathrm{CF} = v) \]
        \item Find the peak index $j_{\max} = \arg\max\left(|\mathbf{z}|\right)$\;
        \item Find the smallest index $i$ such that:
        \[ \mathrm{freq}(i) > v \quad \text{and} \quad \frac{|\mathbf{z}_i|}{|\mathbf{z}_{j_{\max}}|} < \Delta \]
        \item Append $v$ to $\mathrm{CFs}$\;
        \item Update $v$ to the next center frequency:
        \[ v \leftarrow v + \left( \mathrm{freq}(i) - \mathrm{freq}(j_{\max}) \right) \]
    \end{enumerate}
}
\caption{Algorithm to select center frequencies for reduced ripple (CFs). Here, $\mathrm{freq}(i)$ denotes the frequency at index $i$, $\mathrm{frequencyResponse}$ computes the magnitude response of the filter $f^{NH}$ at specified frequencies, and $j_{\max}$ is the index of the peak response.}
\label{algo:CF_spacing}
\end{algorithm}
Fig. \ref{fig:ripple_zilany} shows an example comparison of the conventional log-spacing, used in \cite{Drakopoulos2023,Baby2021AApplications,Nagathil2023WaveNet-basedModel}, and our proposed spacing. 
To better see the effect of our proposed spacing strategy as a function of the number of channels, $K$, we introduce a normalized error, defined as the gain-to-ripple-ratio (GNR):
\begin{equation}
    \textrm{GNR}(\textbf{g}^{ref},\textbf{g}) \stackrel{\text{def}}{=} 10 \log_{10} \left (\dfrac{||\textbf{g}^{ref}||_2^2}{||\textbf{g}^{ref}-\textbf{g}||_2^2} \right) \, ,
\end{equation}
where $\textbf{g}^{ref}$ is the gain of a reference compensation strategy, which can be computed using a large number of CFs (for which the ripple vanishes), and $\textbf{g}$ is the gain of some proposed compensation strategy. In Fig. \ref{fig:spacing}, we compare the GNR of a log spacing and our proposed spacing, using the output of the chirping filter of the Zilany model, see Fig. \ref{fig:zilanyoverview}, which is the primary contributor to the frequency tuning of the model, and  a N3 hearing loss \cite{Bishop2006}. We find that our proposed spacing shows less ripple than the log-spaced CFs. Furthermore, Fig. \ref{fig:spacing} also shows that increasing the number of CFs decreases the ripple of the optimal-compensation gain curve, as  expected.

  \section{DNN-based hearing-loss compensation}
\label{sec:JP2_comp_networks}
 In this section, we present the neural networks and the loss functions used for training the DNN-based HLCs developed in this paper. Additionally, we present an extension to the DNN-HLC that also provides noise reduction. We refer to this network as 'deep-neural-network-based joint hearing-loss compensation and noise reduction' (DNN-HLCNR). 
\subsection{Architecture of the DNN-HLC}
For the DNN-HLC, we used the same neural network as was used for the AME, namely Wave-U-Net \cite{Stoller2018}, with parameters found in Table \ref{tab:DNNHLC}. Wave-U-Net shares its core structure with the DNN used in previous work \cite{Drakopoulos2023ACompensation}, with the key difference being that interpolation is used in the decoder instead of transposed convolution. This change, consistent with findings in prior literature \cite{Stoller2018}, was found to reduce tonal artifacts. The hyperparameters of the Wave-U-Net were found during pilot tests, both by observing convergence of loss functions and by informal listening tests by the authors.
\begin{table}[H]
\centering
      \caption{Hyperparameters of the DNN-HLC. N denotes the number of Upsampling/Downsampling blocks.}
      \label{tab:DNNHLC}
\begin{tabular}{|l|l|}
\hline
N     & 6   \\ \hline
Kernel size                 & 9     \\ \hline
Channels & 128 \\ \hline
Decimation factor & 2 \\ \hline
Encoder activation & Tanh  \\ \hline
Decoder activation & PReLU \\ \hline
Bias & None \\ \hline
\end{tabular}
\end{table}

\subsection{Loss function for DNN-based HLCs}

A commonly used loss function in regression tasks is the mean-absolute error (MAE). We compute the MAE between the normal-hearing and hearing-impaired auditory model as:
\begin{equation} 
    L_{MAE}= \mathbb{E}_\textbf{x}\left[\dfrac{1}{K}\sum_{k=1}^K ||f_k^{NH}(\textbf{x})-f_k^{HI}(c(\textbf{x}))||_1\right]  \, 
    \label{eq:JP2_MAE}
\end{equation}
Such a loss function might be sensitive to minor phase differences between the normal and hearing-impaired auditory model, which may not have perceptual significance. Therefore, we modified the loss function by segmenting the output of each auditory-model channel across the temporal dimension and averaging the response within these segments before computing the difference. In this way, minor phase differences present will not dominate the loss function. We compute the MAE using segments corresponding to different durations: 1 ms, 10 ms, and 100 ms, denoted as $L_{MAE,1}$, $L_{MAE,10}$, and $L_{MAE,100}$, respectively.  While achieving zero MAE at 1 ms ($L_{MAE,1}$) would imply optimal performance across longer time scales, we find both empirically, and for the linear, time-invariant model discussed in Sec. \ref{Sec:opt_comp_strategies}, that complete restoration is infeasible. Incorporating different time scales in the analysis could allow us to account for various auditory effects that are relevant over both shorter and longer durations.

In order to avoid spurious low-frequency behavior caused by the AMEs, which are unable to emulate non-audible low-frequency content, we included a term to penalize low-frequency energy:
\begin{equation}
 L_f = \mathbb{E}_\textbf{x}\left[\sum_i||(\textbf{W}\textbf{x})_i-\textbf{W}(c(\textbf{x}))_i||_1 \right] \, , \, \, \mathrm{freq}(i) < 20 \textrm{Hz} \, , 
    \label{eq:JP2_freq_loss}
\end{equation}
where $\textrm{freq}(i)$ is the frequency of the i-th index of the discrete Fourier transform.
We combined the loss functions as follows: 
\begin{equation}
\label{eq:JP2_composite}
    L = L_{MAE,1} + L_{MAE,10} + L_{MAE,100} + \gamma  L_f
\end{equation}
where $\gamma$ is a small constant, determined empirically through pilot experiments. The exact value of $\gamma$ is not critical due to its negligible impact on the audible output.

\subsection{Architecture of the noise-reduction networks}
\label{sec:JP2_NR_network}
 We trained two different noise-reduction networks: A generic NR network, and our proposed network, the joint DNN-based HLC and NR (DNN-HLCNR). 
 \iffalse : \begin{equation}
   \text{SI-SDR}(\textbf{s},\hat{\textbf{s}}) =  10 \log_{10} \dfrac{||\dfrac{\hat{\textbf{s}}^T{\textbf{s}}}{||\textbf{s}||_2^2}||_2^2}{||\dfrac{\hat{\textbf{s}}^T\textbf{s}}{||\textbf{s}||_2^2}\textbf{s} - \hat{\textbf{s}} ||_2^2}\, ,
   \label{eq:JP2_SISDR}
\end{equation}
where $\textbf{s}$ is the target signal of interest, and $\hat{\textbf{s}}$ is the estimated denoised signal, e.g. the output of a DNN-NR. 
\fi 
 For the two noise-reduction networks, we used identical architectures, a variation of Conv-TasNet \cite{Luo2018Conv-TasNet:Separation}  that is more memory efficient, SuDORMRF \cite{Tzinis2020SudoSeparation}. The neural network consists of an encoder that downsamples a noisy acoustic signal, followed by a separator that calculates a non-negative mask which is applied to the encoded representation of the acoustic signal, resulting in a denoised, encoded representation. Finally, the denoised representation is upsampled by the decoder, resulting in the denoised acoustic signal.  
\iffalse
\begin{figure}
  \centering
      \includegraphics[scale=0.45]{SUDORMf.pdf}
  \caption{Overview of the SUDORmF \cite{Tzinis2020SudoSeparation} architecture used for the two noise reduction networks, the DNN-HLCNR and the generic NR.}
  \label{fig:sudormf}
\end{figure}
\fi

\subsection{Training}

The DNN-HLC and the DNN-HLCNR were designed to perform two different tasks, and were therefore trained in two different ways:
\subsubsection{DNN-HLC}
In previous work \cite{Drakopoulos2023a}, the DNN-HLC was trained using clean speech but tested on noisy speech. However, we found that this approach could lead to undesirable distortions during testing, likely due to a lack of generalization to noisy conditions. To mitigate this, we trained our proposed DNN-HLC using noisy speech inputs to ensure it could robustly handle noisy environments during testing. Specifically, in Eqs. \ref{eq:JP2_MAE}
-\ref{eq:JP2_composite}, the input signal, $\textbf{x}$, was modeled as a combination of clean speech, $\textbf{s}$, and additive noise, $\textbf{v}$, such that $\textbf{x} = \textbf{s} + \textbf{v}$, exposing the DNN to realistic conditions during training.
\subsubsection{DNN-HLCNR}
In order to train a joint DNN-HLC and NR (DNN-HLCNR), we trained a noise-reduction network, cf. Sec. \ref{sec:JP2_NR_network}. To do so,  $f^{NH}(\textbf{x})$ was replaced with $f^{NH}(\textbf{s})$ in Eq. (\ref{eq:JP2_MAE}) and $\textbf{W}\textbf{x}$ was replaced with $\textbf{W}\textbf{s}$ in Eq. (\ref{eq:JP2_freq_loss}). The noisy mixture, $\textbf{x}$, was used as input to the DNN-HLCNR.

\subsection{Training methodology and data}
All DNNs were trained using 8000 utterances for training and 2000 utterances for validation. All utterances were sampled from LibriTTS \cite{Zen2019LibriTTS:Text-to-Speech} and were not seen during training the AMEs, cf. Sec. \ref{sec:JP2_AME}. The DNNs were trained, using a learning rate of 0.0001 with an ADAM optimizer for maximally 20 epochs. Training was often terminated before (or at) the 20th epoch by using early-stopping, evaluating the loss function on the validation set. Additionally, the gradients were clipped to avoid exploding gradients that could arise from the idiosyncrasies of the AMEs. All utterances were normalized to a SPL of 65 dB, corresponding to normal conversation levels, at which the model is tested. Two different noise types were used: 16-speaker babble noise and stationary, Gaussian, speech-shaped noise. The SNRs were randomly selected from a wide range of SNRs that were relevant for the listening experiments, specifically $\{ 0, 3, 6, 9, 12, \text{clean}\}$ dB SNR. The inclusion of the "clean" SNR, represents near-ideal, noise-free conditions, allowing the network to retain high performance even in the absence of noise.
One DNN-HLC and one DNN-HLCNR were trained for each test person (TP), i.e. for each individual audiogram. A single, generic noise-reduction network, was trained using SI-SDR \cite{Roux2018SDRDone} as a loss function, to serve as a universal noise reduction network that could be applied across different listeners.

\subsection{Pilot tests}
\label{sec:JP2_pilot}
\begin{figure}[]
  \centering
      \includegraphics[scale=0.5]{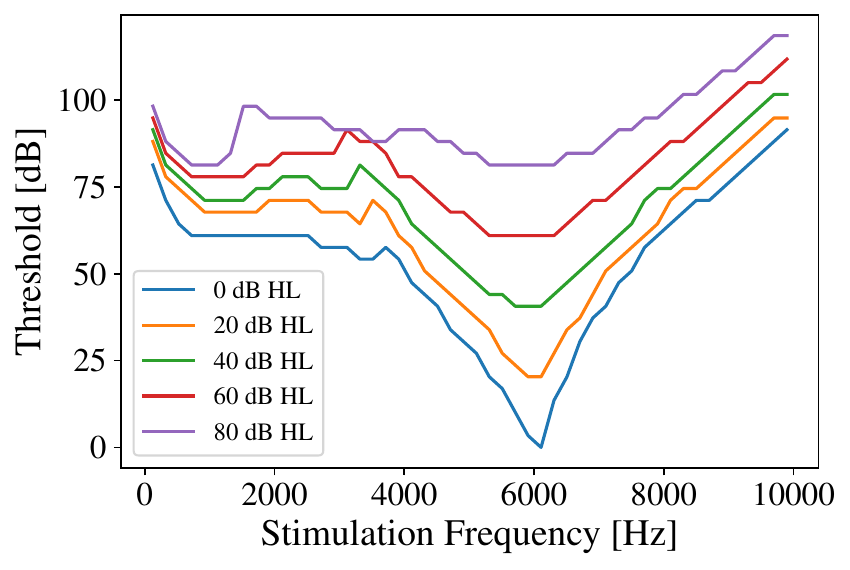}
  \caption{Thresholds at the auditory nerve level for different levels of hearing loss, using the Zilany auditory model, at a CF of 6 kHz using tonal stimuli. The thresholds are calculated, for each stimulation frequency, as the level where the mean firing rate is 1 standard deviation larger than spontaneous firing rate.}
  \label{fig:thresholds}
\end{figure}

\begin{figure}[]
  \centering
      \includegraphics[scale=0.5]{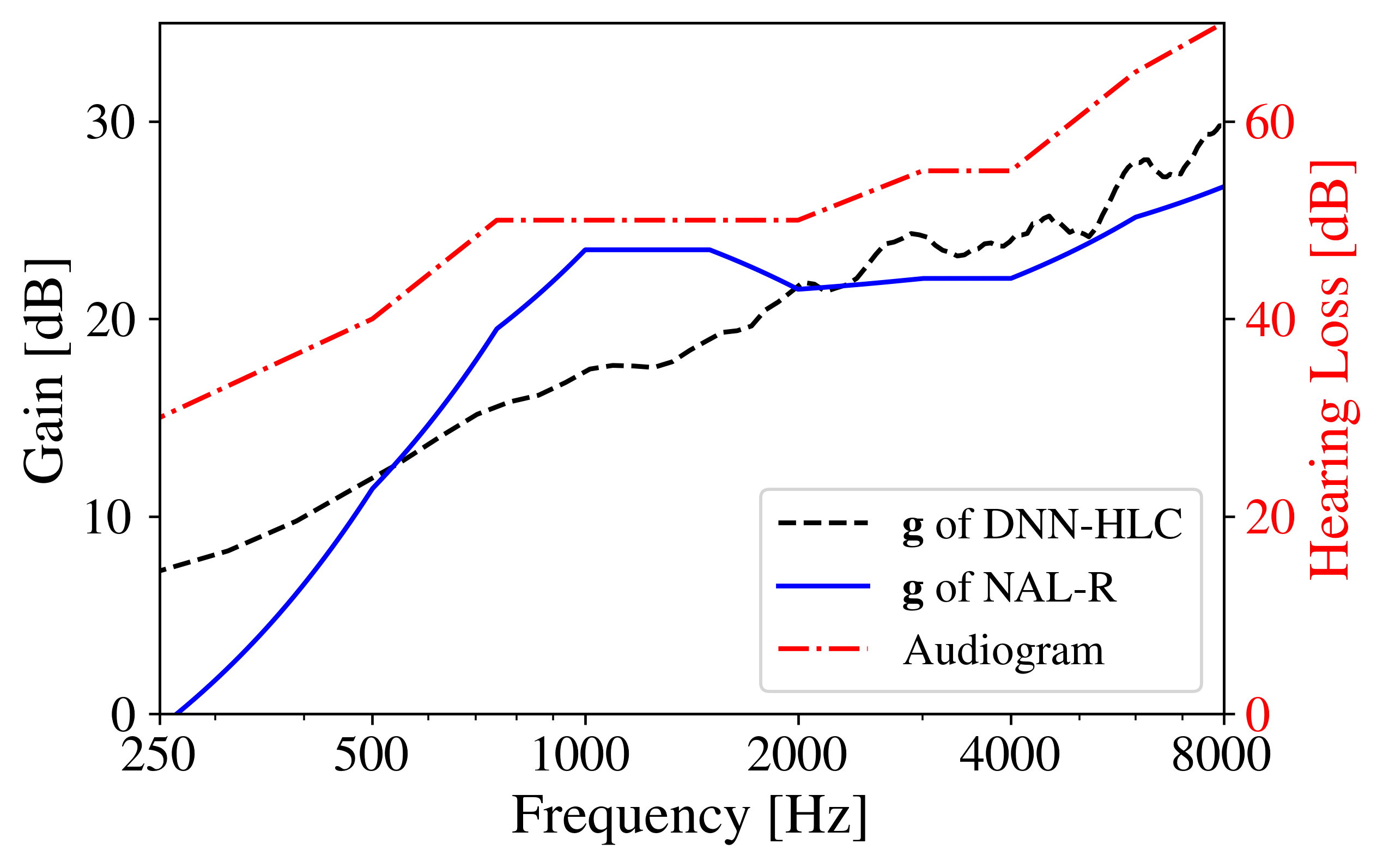}
  \caption{Comparison of $\textbf{g}$ for the linear compensation strategy, NAL-R, and the DNN-based hearing-loss compensation strategy, DNN-HLC for a given test person, with the audiogram shown in red.}
  \label{fig:gain_example}
\end{figure}

During the first pilot tests with hearing-impaired test persons (TPs) we found that the DNN-HLC and DNN-HLCNR did not perform very well, particularly at high frequencies, where large tonal components were introduced. Additionally, excessive low-frequency gain was provided by the DNNs. Our hypothesis is that this behavior was caused by the frequency response of the auditory-model channels essentially becoming flat for sufficiently large hearing losses, see Fig. \ref{fig:thresholds}.. We conjecture this behavior is caused by the flat filters at high CFs, since for a sufficiently broad tuned auditory nerve, any frequency can be used to drive high firing rates for a given auditory model channel. In order to circumvent this problem, we experimented with different loss functions, attempting to impose spectral smoothness on the long-term gain, but this did not significantly alleviate the issue. Additionally, we appended a model of the inferior colliculus (IC) to the output of the auditory nerve, which essentially works as a non-linear modulation filter bank \cite{Carney2015}. Appending the model of the IC reduced tonal components, as tonal components could not be used to drive modulation, but still did not fix the issue of elevated low-frequency content.  Thus, we instead opted to not include the IC model in the final listening tests and reduce the hearing loss in the auditory model, by dividing the audiogram by a factor 2, inspired by the half-gain rule \cite{Lybarger1944}. This procedure directly imposes a constraint on the firing rate being driven by frequencies around the CF of each channel. Additionally, we found that the trained DNNs resulted in gain curves within the same order of magnitude as NAL-R. A comparison of the resulting gain, $\textbf{g}$, of NAL-R and DNN-HLC is shown for a realistic hearing loss in Fig. \ref{fig:gain_example}.

  \section{Experiments}
\label{sec:JP2_experiments}
 In this section we describe the two listening tests we conducted: the Danish Hearing in Noise Test (HINT) \cite{Nielsen2011TheTest} with the goal of testing speech intelligibility at difficult SNRs using different compensation strategies, and 2) a  Multiple Stimuli with Anchor (MUSA) test, inspired by the MUSHRA test \cite{RadiocommunicationSector2015Methodsound}, with the goal of quantifying which compensation strategies the test persons (TPs) preferred.
\subsection{Test persons}
\begin{figure}[!h]
  \centering
      \includegraphics[scale=0.5]{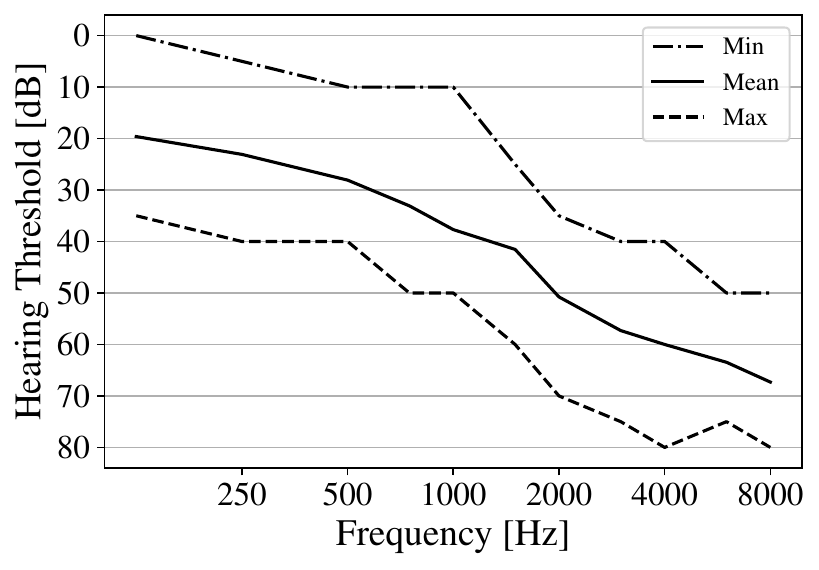}
  \caption{Distribution of the left-right-ear averaged pure-tone audiograms of all test persons (TPs). Min denotes the minimal hearing threshold of the TPs, and max denotes the maximum hearing threshold of the TPs. Mean denotes the average hearing threshold of the 13 TPs.}
  \label{fig:audiogram}
\end{figure}

Thirty TPs were initially contacted for experiments. The TPs were restricted to having relatively symmetric hearing losses ($\pm 5$ dB) between the two ears. Twenty TPs accepted the invitation, and the first five of these participated in pilot tests. Out of the remaining 15 TPs,  two were excluded due to their inability to complete the tests.  
The statistics of the left-and-right averaged audiograms of the TPs that were used for the final listening experiments, excluding the pilot tests, cf. Sec \ref{sec:JP2_pilot}, are shown in Fig. \ref{fig:audiogram}. 
The average age of the 13 TPs used for the final listening experiments was 75.5 years, of which eight TPs were biologically male and five were biologically female.
\subsection{Testing conditions}
The listeners were placed in a soundproof booth, wearing Sennheiser HDA200 headphones. Each signal was always normalized at peak magnitude after offline processing by each strategy. This normalization partially accounted for loudness differences between strategies, while ensuring that no clipping occurred. The volume was personally adjusted by the individual, and the level was tested for the various strategies to guarantee that the signals were loud enough to be heard. We ensured that the signals surpassed the hearing threshold by: 1) assessing their listening skills on the strategies using HINT lists not utilized during data collection, and 2) instructing the TP to make the signals "slightly" loud, gradually increasing the level until the signal were well above the threshold. The volume stayed fixed during all of the tests.

In our comparison between DNN-HLC and a conventional HLC, we chose to employ the NAL-R fitting rationale as the conventional HLC. This decision was guided by NAL-R \cite{Byrne1986TheAid} being linear and non-proprietary, making the results easier replicable. While newer wide dynamic range based strategies like NAL-NL2 \cite{Keidser2011} are often used in hearing aids to deal with the large dynamic range of everyday sounds, it's important to note that our study focused on a fixed SPL level. The primary advantage of utilizing the NAL-NL2 fitting rationale lies in its capability to span a broad range of input SPLs. Under conditions with sufficiently low compression ratios, relatively stationary noise and longer time constants, NAL-NL2 may essentially function as a linear filter at a fixed SPL, resulting in an essentially linear strategy. Additionally, we compare our joint DNN-HLCNR with a generic DNN-NR, followed by NAL-R, cf. Sec. \ref{sec:JP2_comp_networks}.
The five conditions are as follows:
\begin{enumerate}
    \item Unprocessed (flat gain, individually adjusted)
    \item NAL-R  \cite{Byrne1986TheAid} (prescribed linear gain)
    \item DNN-HLC 
    \item Generic NR followed by NAL-R (defined as NR + NAL-R)
    \item DNN-HLCNR  (joint HLC and NR)
\end{enumerate}
\subsection{Speech intelligibility test}
For the HINT, we wanted to target the expected steepest part of the psychometric function, i.e. $50\%$ word score, and therefore generated signals at 0 dB and 3 dB SNR  \cite{Nielsen2009} with 16-speaker babble noise. The signals were then processed by all the strategies, totaling 10 different conditions. For each subject, each condition was assigned to a random HINT list, of which there were 10 ($2$ SNRs $\times $ $5$ strategies). Each HINT list consists of 20 sentences with 5 word in each sentence, and the subjects were instructed to repeat the words they heard. If the subjects were unable to discern any words, they were instructed to say "pass". Additionally, the subjects were instructed that guessing was allowed if they were uncertain of some specific word. The answers were recorded and scored by the first author and a clinical audiologist following standard clinical practice.

\subsection{Preference test}
For the MUSA test, the signals were generated at a SNR of 9 dB --- close to perfect intelligibility ---, assuming a perceived sound quality degeneration from the noise. Instead of including the unprocessed condition at 9 dB SNR, we included a hidden anchor, which was the unprocessed signal at an SNR of 3 dB. The anchor, firstly, provided a lower fixed-point for the 0-100 MUSA scale. Secondly, the anchor could be used for detecting subjects that rated the anchor favorably, thus not performing the tests properly. Naturally, as we did not have access to a ground truth reference, i.e. a processing of signal that is truly optimal for each individual, for each sentence, we did not include a reference, hence the MUSA acronym. The test signals were the first 15 sentences of the Danish HINT, List 1. For each sentence, each strategy was given a random number (1 through 5) and the subjects were instructed to rank the strategies from the  the least preferred (0) to the most preferred (100). The subject performed a ranking using a computer interface with 5 sliders, and a play button for each number (1 through 5). The MUSA test was self-paced, allowing subjects to listen to the five strategies as many times as needed before making their ratings.

\section{Results}
\label{sec:JP2_results}
In this section we present our results from the listening tests, where we measure the performance of our proposed hearing-loss compensation strategies.
\subsection{HINT results}
\begin{figure*}[!h]
  \centering
      \includegraphics[scale=0.5]{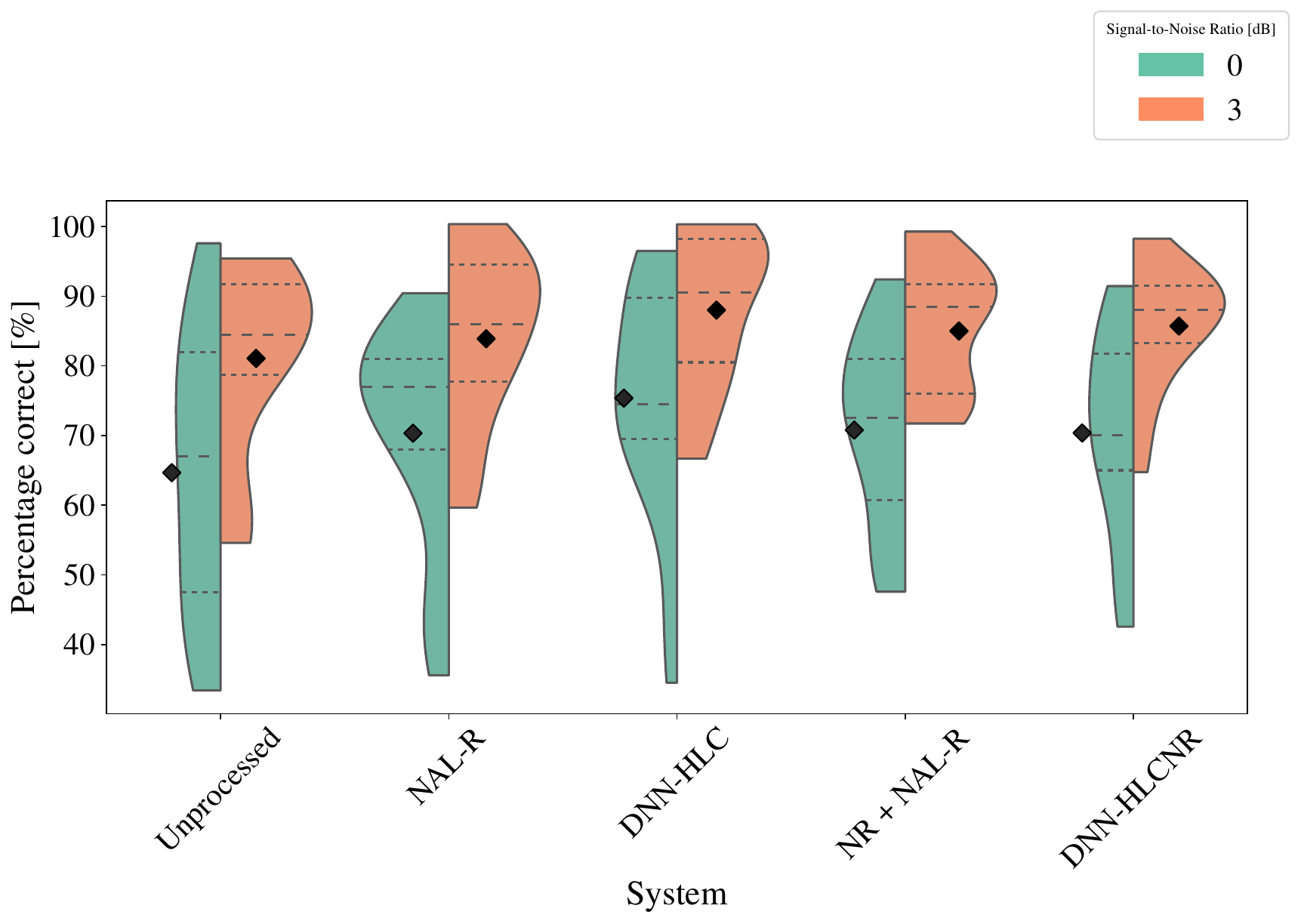}
  \caption{Violin plot of the results for the hearing in noise test (HINT) for the 5 different systems at two different SNRs. The black diamonds denote the mean percentage correct of each condition, and the inner lines are the 1st, 2nd and 3rd quantile.}
  \label{fig:HINT_boxplot}
\end{figure*}

 Fig. \ref{fig:HINT_boxplot} shows the results of the HINT, displayed as a violinplot. By performing a multiple comparison of means using a Tukey's range test, we found no significant difference between any of the conditions, using a family-wise error rate (FWER) of $\alpha=0.05$. Instead, we fit the data using a linear mixed model (LMM) \cite{Raudenbush2001} after transforming the data using a rationalized arcsine transform to stabilize variances and handle floor and ceiling effects \cite{Studebaker1985ATransform}. The LMM models the random effect of different TPs, and the additive, fixed effect of SNR, using the following formula:
\begin{align}
\text{Intelligibility} &\sim C(\text{System, Treatment('Unprocessed')}) \nonumber \\ &+ C(\text{SNR}) + \text{Random Effects of TP} + \epsilon
\end{align}
where $\text{C}(\cdot)$ denotes a categorical variable, the Treatment refers to setting the mean of the "Unprocessed" system as the intercept, and $\epsilon$ as the residuals. After fitting the model, the assumption of normality of the residuals could not be rejected by using Shapiro-Wilk \cite{SHAPIRO1965AnSamples}, Anderson-Darling \cite{Anderson1954AFit} and Jarque-Bera tests \cite{Jarque1987AResiduals}.

The results of such a LMM, shown in Table \ref{tab:LMM}, indicates that there is a significant improvement ($p \approx 0.00007$) using the DNN-HLC, leading to a $13$ (RAU units) increased intelligibility, as compared to the unprocessed system. Similarly, there is significant evidence ($p\approx 0.04$) that the NR-based systems also provide increased intelligibly as compared to the unprocessed baseline. 

\begin{table}[H]
\caption{Coefficients of a Linear Mixed Model fitted to the outcome of the HINT test. The p-values are adjusted using Benjamini-Hochberg correction, and $\epsilon$ denotes the residuals.}
\label{tab:LMM}
\centering
\begin{tabular}{|l|l|l|}
\hline
Variable                & Coeff. & Adj. p-value \\ \hline
Intercept (Unprocessed) & 62          & 0              \\ \hline
C(NAL-R)                & 6           & 0.078        \\ \hline
C(DNN-HLC)              & 13           & 0.00007    \\ \hline
C(NR+NAL-R)             & 7           & 0.04         \\ \hline
C(DNN-HLCNR)            & 7           & 0.04          \\ \hline
C(SNR=3)                & 20          & 0      \\ \hline
Variance of Random effects of TP + $\epsilon$               & 158          & 0.039   \\ \hline
\end{tabular}
\end{table}

\subsubsection{Pairwise comparisons}

\hl{Two planned pairwise comparisons were performed using contrast testing within the LMM framework, corresponding to systems without and with NR:
\begin{enumerate}
    \item  \textbf{NAL-R vs. DNN-HLC}: To evaluate the improvement in intelligibility when using the DNN-HLC system compared to standard NAL-R processing.
    \item \textbf{DNN-HLCNR vs. 'NR + NAL-R'}: To determine if the DNN-HLCNR system offers a significant advantage over the combination of a generic NR system and NAL-R.
\end{enumerate}

Contrast vectors were constructed to represent these two hypotheses. P-values were adjusted for multiple comparisons using the Benjamini-Hochberg procedure. The comparison between NAL-R and DNN-HLC showed a significant improvement in intelligibility scores with the DNN-HLC system, with an adjusted p-value of $p = 0.015$. In contrast, the comparison between DNN-HLCNR and 'NR + NAL-R' did not suggest a significant difference in intelligibility scores. 
}
\subsubsection{Comparison with HASPI}

We also investigated the relationship between HASPI \cite{Kates2021The2} scores and the measured intelligibility. Using Spearman's rank correlation we found a correlation coefficient of 0.10 with a p-value of 0.80  and using Kendall's rank correlation coefficient, we found a correlation coefficient of 0.07 with a p-value of 0.70. Thus, we could establish no correlation between HASPI and the intelligibility results. This result is consistent with previous studies on DNN-based speech enhancement in normal-hearing listeners \cite{Lopez-Espejo2023OnIntelligibility}, which reported weak correlation between subjective listening tests and conventional objective metrics.
\subsection{MUSA results}
\begin{figure*}[!h]
  \centering      \includegraphics[scale=0.5]{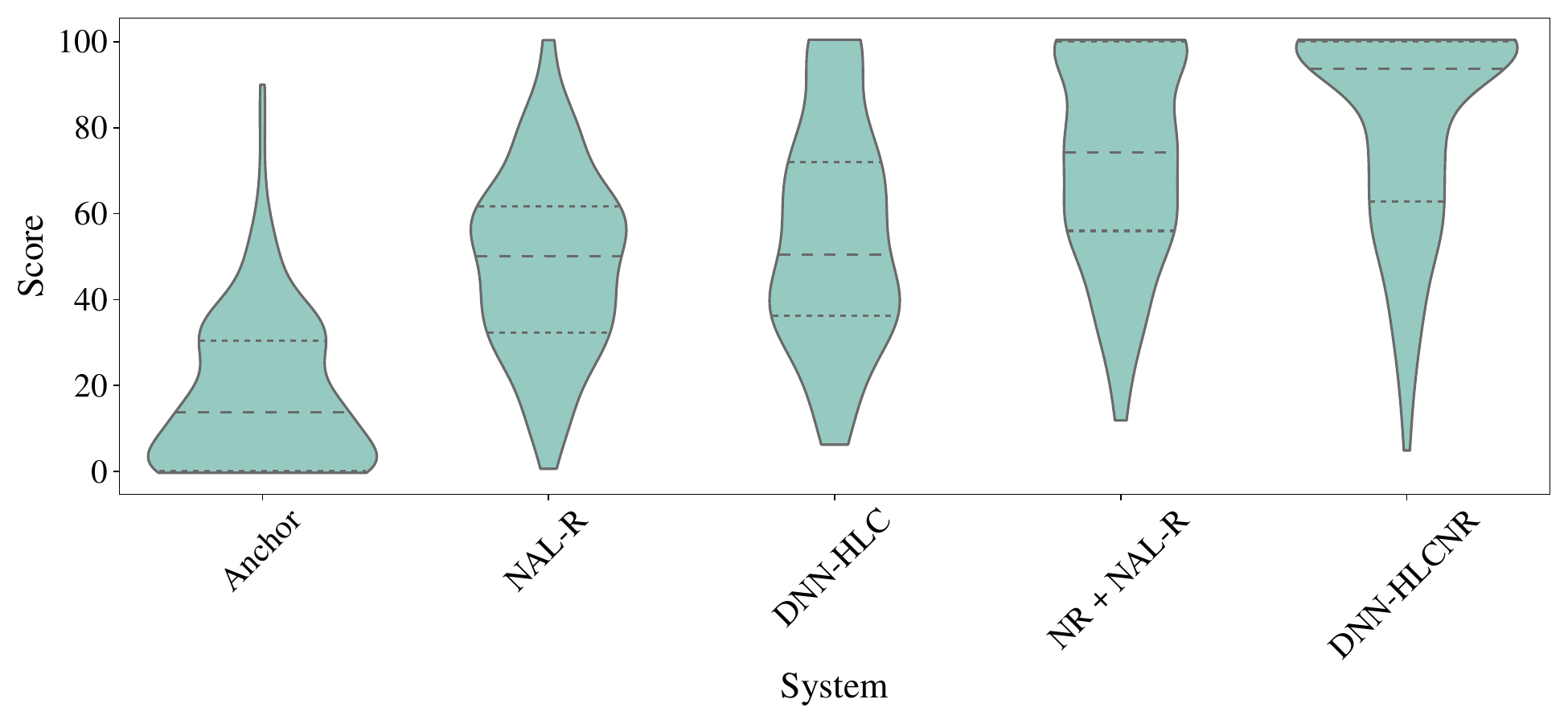}
  \caption{Distribution of the results of the preference (MUSA) test for 5 different conditions (Anchor, NAL-R, DNN-HLC, NR + NAL-R, DNN-HLCNR), indicating different processing strategies.}
  \label{fig:MUSHRA_distribution}
\end{figure*}

Fig. \ref{fig:MUSHRA_distribution} shows a violin plot of the data collected from the MUSA test. A Wilcoxon signed-rank test was performed between all the different systems. All possible comparisons were significant using a FWER of $\alpha=0.05$ after using Benjamini-Hochberg corrections. The average and median score of each system are shown in Table \ref{tab:results}, where we find that DNN-HLC is preferred over NAL-R by a small margin, and that the DNN-HLCNR is the most preferred system, with a median score of 94:
\begin{table}[!h]
\centering
\caption{Average and median score of the MUSA test for each system}
\label{tab:results}
\begin{tabular}{|l|l|l|}
\hline
System    & Mean score & Median score \\ \hline
Anchor    & 18         & 14           \\ \hline
NAL-R     & 49         & 50           \\ \hline
DNN-HLC    & 55         & 51           \\ \hline
NR + NAL-R & 72         & 74           \\ \hline
DNN-HLCNR & 81         & 94           \\ \hline
\end{tabular}

\end{table}

  \section{Discussion}
\label{sec:JP2_Discussion}
In the HINT, the intelligibility test, we found that the DNN-based hearing-loss compensation strategies (DNN-HLC) performed comparably to the joint hearing-loss compensation and noise-reduction strategies (DNN-HLCNR). It is somewhat counterintuitive that the DNN-HLCNR does not significantly outperform the DNN-HLC, as we would expect NR to increase intelligibility. One reason might be that noise lreduction is known to sometimes generalize poorly to out-of-distribution datasets, particularly when the out-of-distribution dataset does not contain the same speakers as the original training set. \cite{Gonzalez2023AssessingEnvironments}. In this work, the DNN-HLCNR was trained on English speakers, meanwhile our test data was from different speakers in Danish, which could be problematic. In the MUSA test, that measured preference, the DNN-HLCNR outperformed all other systems, which could be due to a more appropriate compensation for the individual. Alternatively this difference could be due to better generalization capabilities of the DNN-HLCNR or the increased low-frequency content of the DNN-HLCNR, which is certainly not expected to reduce intelligibility in high SNRs.

Unfortunately, we found no evidence of correlation between HASPI and the intelligibility scores in the listening experiments: This makes it difficult to conduct any ablation experiments and disprove or verify conjectures related to improvements or discrepancies in performance. While we know that our ultimate goals are improved intelligibility and speech quality, there is no straightforward mapping from peripheral auditory processing to these high-level percepts. This points to a fundamental problem with the approach suggested in this work: Choosing a suitable loss function and measuring its effectiveness are two sides of the same coin. On one hand, without a clear understanding of how to connect peripheral changes to perceptual outcomes, it’s challenging to determine what loss function best aligns with these goals. On the other hand, without precise, effective and reliable ways to measure intelligibility and quality improvements, evaluating the effectiveness of any chosen loss function becomes equally problematic. As a result, it becomes difficult to iterate quickly or draw firm conclusions about performance, especially without extensive, resource-heavy testing with hearing-impaired individuals. 

In this work, we have demonstrated that using established principles from signal processing and deep learning can lead to viable strategies for hearing-loss compensation. However, these approaches may still be suboptimal and require further refinement. Specifically, the importance of temporal non-linear processing in the DNN-HLC, and how it potentially diverges from traditional methods compared to its long-term frequency shaping characteristics, warrants deeper investigation in future studies. Additionally, while we fixed the input sound level at 65 dB SPL (normal conversational levels) in this work, real-world environments feature a much broader range of sound intensities. Thus, it remains an open question how to ensure that the DNN-HLC can effectively process a wide range of input levels while ensuring a safe and comfortable listening experience for the hearing-impaired. Future research should also investigate how different loss functions interact with specific types of hearing loss and the implications these interactions have on final compensation strategies.

  \section{Conclusion}
\label{sec:JP2_conclusion}
In this work we presented a deep-neural-network (DNN) based framework for providing personalized hearing loss compensation (HLC) and noise reduction (NC) strategies using auditory models. An auditory model of the auditory pathway, including an auditory-nerve model was emulated using a DNN, which subsequently was used to train the HLC and NR signal-processing strategies. The HLC strategy was analyzed using a simple, linearized model. The analysis showed that some of the properties of the DNN-based hearing-loss compensation strategy could be predicted before training the DNN, which could assist in hyperparameter choices of the auditory model, e.g. spacing of center frequencies of the auditory model.
Finally, the derived HLC and NR strategies were evaluated in terms of intelligibility and preference and compared to conventional signal-processing strategies. In terms of intelligibility in a difficult listening scenario, the listening experiments showed that the resulting DNN-based HLC and NR strategies outperformed a baseline consisting of unprocessed signals. \hl{Additionally, we found that the proposed DNN-based HLC outperformed the conventional NAL-R strategy.} In terms of preference, the proposed joint HLC and NR outperformed a generic NR followed by a conventional HLC strategy. Although the results are promising, several issues remain to be addressed, such as generalization to unseen types of noise, languages and speakers, characterization and explanation of the DNN-based hearing-loss compensation strategies and insuring stability and performance across a wide range of inputs.

\newpage
\bibliographystyle{IEEEtran}
\bibliography{IEEEabrv,references}

\begin{thebibliography}{10}
\providecommand{\url}[1]{#1}
\csname url@samestyle\endcsname
\providecommand{\newblock}{\relax}
\providecommand{\bibinfo}[2]{#2}
\providecommand{\BIBentrySTDinterwordspacing}{\spaceskip=0pt\relax}
\providecommand{\BIBentryALTinterwordstretchfactor}{4}
\providecommand{\BIBentryALTinterwordspacing}{\spaceskip=\fontdimen2\font plus
\BIBentryALTinterwordstretchfactor\fontdimen3\font minus
  \fontdimen4\font\relax}
\providecommand{\BIBforeignlanguage}[2]{{%
\expandafter\ifx\csname l@#1\endcsname\relax
\typeout{** WARNING: IEEEtran.bst: No hyphenation pattern has been}%
\typeout{** loaded for the language `#1'. Using the pattern for}%
\typeout{** the default language instead.}%
\else
\language=\csname l@#1\endcsname
\fi
#2}}
\providecommand{\BIBdecl}{\relax}
\BIBdecl

\bibitem{Rasiah2018AddressingLoss}
{Rasiah} and {Sulakshan}, ``{Addressing the rising prevalence of hearing
  loss},'' \emph{World Health Organization}, 2018.

\bibitem{Akeroyd2020LaunchingProcessing}
\BIBentryALTinterwordspacing
M.~A. Akeroyd, J.~P. Barker, T.~J. Cox, J.~Culling, S.~Graetzer, G.~Naylor,
  E.~Porter, and R.~Viveros~Mu{\~{n}}oz, ``{Launching the first “Clarity”
  Machine Learning Challenge to revolutionise hearing device processing},''
  \emph{The Journal of the Acoustical Society of America}, vol. 148, no.
  4{\_}Supplement, pp. 2711--2711, 10 2020. [Online]. Available:
  \url{/asa/jasa/article/148/4_Supplement/2711/645306/Launching-the-first-Clarity-Machine-Learning}
\BIBentrySTDinterwordspacing

\bibitem{Akeroyd2023ResultsDevices}
\BIBentryALTinterwordspacing
M.~A. Akeroyd, J.~L. Firth, G.~Naylor, J.~P. Barker, J.~Culling, T.~J. Cox,
  W.~Bailey, S.~Graetzer, R.~Viveros~Mu{\~{n}}oz, E.~Porter, and H.~Griffiths,
  ``{Results of the second “clarity” enhancement challenge for hearing
  devices},'' \emph{The Journal of the Acoustical Society of America}, vol.
  153, no. 3{\_}supplement, pp. A48--A48, 3 2023. [Online]. Available:
  \url{https://www.researchgate.net/publication/370320818_Results_of_the_second_clarity_enhancement_challenge_for_hearing_devices}
\BIBentrySTDinterwordspacing

\bibitem{Cox2023OverviewAids}
\BIBentryALTinterwordspacing
T.~J. Cox, J.~Barker, W.~Bailey, S.~Graetzer, M.~A. Akeroyd, J.~F. Culling, and
  G.~Naylor, ``{Overview Of The 2023 Icassp Sp Clarity Challenge: Speech
  Enhancement For Hearing Aids},'' \emph{ICASSP, IEEE International Conference
  on Acoustics, Speech and Signal Processing - Proceedings}, 11 2023. [Online].
  Available: \url{https://arxiv.org/abs/2311.14490v1}
\BIBentrySTDinterwordspacing

\bibitem{Nagathil2021ComputationallyProcessing}
A.~Nagathil, F.~G{\"{o}}bel, A.~Nelus, and I.~C. Bruce, ``{Computationally
  efficient DNN-based approximation of an auditory model for applications in
  speech processing},'' \emph{ICASSP, IEEE International Conference on
  Acoustics, Speech and Signal Processing - Proceedings}, vol. 2021-June, pp.
  301--305, 2021.

\bibitem{Nagathil2023WaveNet-basedModel}
A.~Nagathil and I.~C. Bruce, ``{WaveNet-based approximation of a cochlear
  filtering and hair cell transduction model},'' \emph{The Journal of the
  Acoustical Society of America}, vol. 154, no.~1, pp. 191--202, 7 2023.

\bibitem{Baby2021AApplications}
\BIBentryALTinterwordspacing
D.~Baby, A.~V.~D. Broucke, and S.~Verhulst, ``{A convolutional neural-network
  model of human cochlear mechanics and filter tuning for real-time
  applications},'' \emph{Nature machine intelligence}, vol.~3, no.~2, p. 134, 2
  2021. [Online]. Available:
  \url{https://www.ncbi.nlm.nih.gov/pmc/articles/PMC7116797/}
\BIBentrySTDinterwordspacing

\bibitem{Drakopoulos2021ASynapses}
\BIBentryALTinterwordspacing
F.~Drakopoulos, D.~Baby, and S.~Verhulst, ``{A convolutional neural-network
  framework for modelling auditory sensory cells and synapses},''
  \emph{Communications biology}, vol.~4, no.~1, 12 2021. [Online]. Available:
  \url{https://pubmed.ncbi.nlm.nih.gov/34211095/}
\BIBentrySTDinterwordspacing

\bibitem{Biondi1978AuditoryViewpoint}
E.~Biondi, ``{Auditory processing of speech and its implications with respect
  to prosthetic rehabilitation. The bioengineering viewpoint},''
  \emph{International Journal of Audiology}, vol.~17, no.~1, pp. 43--50, 1978.

\bibitem{Bondy2004}
J.~Bondy, S.~Becker, I.~Bruce, L.~Trainor, and S.~Haykin, ``{A novel
  signal-processing strategy for hearing-aid design: Neurocompensation},''
  \emph{Signal Processing}, vol.~84, no.~7, pp. 1239--1253, 2004.

\bibitem{Chen2005AMethod}
\BIBentryALTinterwordspacing
Z.~Chen, S.~Becker, J.~Bondy, I.~C. Bruce, and S.~Haykin, ``{A Novel
  Model-Based Hearing Compensation Design Using a Gradient-Free Optimization
  Method},'' \emph{Neural Computation}, vol.~17, no.~12, pp. 2648--2671, 12
  2005. [Online]. Available:
  \url{https://dx.doi.org/10.1162/089976605774320575}
\BIBentrySTDinterwordspacing

\bibitem{Hengel2015SimulatingAids}
\BIBentryALTinterwordspacing
P.~v. Hengel, ``{Simulating hearing loss with a transmission-line model for the
  optimization of hearing aids},'' \emph{Proceedings of the International
  Symposium on Auditory and Audiological Research}, vol.~5, pp. 181--188, 12
  2015. [Online]. Available:
  \url{https://proceedings.isaar.eu/index.php/isaarproc/article/view/2015-21}
\BIBentrySTDinterwordspacing

\bibitem{Drakopoulos2023a}
\BIBentryALTinterwordspacing
F.~Drakopoulos and S.~Verhulst, ``{A Neural-Network Framework for the Design of
  Individualised Hearing-Loss Compensation},'' \emph{IEEE/ACM Transactions on
  Audio, Speech, and Language Processing}, vol.~31, pp. 2395--2409, 7 2023.
  [Online]. Available: \url{http://arxiv.org/abs/2207.07091
  https://ieeexplore.ieee.org/document/10141861/}
\BIBentrySTDinterwordspacing

\bibitem{Verhulst2018}
S.~Verhulst, A.~Alto{\`{e}}, and V.~Vasilkov, ``{Computational modeling of the
  human auditory periphery: Auditory-nerve responses, evoked potentials and
  hearing loss},'' \emph{Hearing Research}, vol. 360, pp. 55--75, 2018.

\bibitem{Taal2011AnSpeech}
C.~H. Taal, R.~C. Hendriks, R.~Heusdens, and J.~Jensen, ``{An algorithm for
  intelligibility prediction of time-frequency weighted noisy speech},''
  \emph{IEEE Transactions on Audio, Speech and Language Processing}, vol.~19,
  no.~7, pp. 2125--2136, 2011.

\bibitem{Kates2021The2}
J.~M. Kates and K.~H. Arehart, ``{The Hearing-Aid Speech Perception Index
  (HASPI) Version 2},'' \emph{Speech Communication}, vol. 131, pp. 35--46, 7
  2021.

\bibitem{Drgas2024DynamicCompensation}
S.~Drgas, L.~Bramslow, A.~Politis, G.~Naithani, and T.~Virtanen, ``{Dynamic
  Processing Neural Network Architecture for Hearing Loss Compensation},''
  \emph{IEEE/ACM Transactions on Audio Speech and Language Processing},
  vol.~32, pp. 203--214, 2024.

\bibitem{LeerEtAl1}
\BIBentryALTinterwordspacing
P.~Leer, J.~Jensen, Z.-H. Tan, J.~{\O}stergaard, and L.~Bramsl{\o}w, ``{How to
  Train Your Ears: Auditory-Model Emulation for Large-Dynamic-Range Inputs and
  Mild-to-Severe Hearing Losses},'' \emph{IEEE/ACM Transactions on Audio,
  Speech, and Language Processing}, vol.~32, pp. 2006--2020, 2024. [Online].
  Available: \url{https://ieeexplore.ieee.org/document/10473115/}
\BIBentrySTDinterwordspacing

\bibitem{K.V.Lindley2002}
\BIBentryALTinterwordspacing
C.~V. Palmer and G.~A. Lindley~Iv, ``{Overview And Rationale For Prescriptive
  Formulas for Linear and Non-Linear Hearing Aids},'' in \emph{Strategies for
  Selecting and Verifying Hearing Aid Fittings.}, P.~Michael~Valente, Ed.\hskip
  1em plus 0.5em minus 0.4em\relax Thieme Medical Publishers, 2002, pp. 1--22.
  [Online]. Available:
  \url{https://web.thieme.com/media/samples/pubid1013629716/}
\BIBentrySTDinterwordspacing

\bibitem{Byrne1986TheAid}
\BIBentryALTinterwordspacing
D.~Byrne and H.~Dillon, ``{The National Acoustic Laboratories' (NAL) new
  procedure for selecting the gain and frequency response of a hearing aid},''
  \emph{Ear and hearing}, vol.~7, no.~4, pp. 257--265, 1986. [Online].
  Available: \url{https://pubmed.ncbi.nlm.nih.gov/3743918/}
\BIBentrySTDinterwordspacing

\bibitem{LindleyIV1997FittingAids}
\BIBentryALTinterwordspacing
G.~A. Lindley~IV and C.~V. Palmer, ``{Fitting Wide Dynamic Range Compression
  Hearing Aids},'' \emph{American Journal of Audiology}, vol.~6, no.~3, pp.
  19--28, 1997. [Online]. Available:
  \url{https://pubs.asha.org/doi/10.1044/1059-0889.0603.19}
\BIBentrySTDinterwordspacing

\bibitem{Schneider1997MultichannelAid}
T.~Schneider and R.~Brennan, ``{Multichannel compression strategy for a digital
  hearing aid},'' \emph{ICASSP, IEEE International Conference on Acoustics,
  Speech and Signal Processing - Proceedings}, vol.~1, pp. 411--414, 1997.

\bibitem{Kates2005MultichannelWarping}
\BIBentryALTinterwordspacing
J.~M. Kates and K.~H. Arehart, ``{Multichannel dynamic-range compression using
  digital frequency warping},'' \emph{Eurasip Journal on Applied Signal
  Processing}, vol. 2005, no.~18, pp. 3003--3014, 2005. [Online]. Available:
  \url{https://www.researchgate.net/publication/26532062_Multichannel_Dynamic-Range_Compression_Using_Digital_Frequency_Warping}
\BIBentrySTDinterwordspacing

\bibitem{Zilany2014}
M.~S.~A. Zilany, I.~C. Bruce, and L.~H. Carney, ``{Updated parameters and
  expanded simulation options for a model of the auditory periphery},''
  \emph{The Journal of the Acoustical Society of America}, 2014.

\bibitem{AuditoryCenter}
\BIBentryALTinterwordspacing
``{Auditory Models - Publications - Carney Lab - University of Rochester
  Medical Center}.'' [Online]. Available:
  \url{https://www.urmc.rochester.edu/labs/carney/publications-code/auditory-models.aspx}
\BIBentrySTDinterwordspacing

\bibitem{Zilany2009ADynamics}
\BIBentryALTinterwordspacing
M.~S.~A. Zilany, I.~C. Bruce, P.~C. Nelson, and L.~H. Carney, ``{A
  phenomenological model of the synapse between the inner hair cell and
  auditory nerve: Long-term adaptation with power-law dynamics},'' \emph{The
  Journal of the Acoustical Society of America}, vol. 126, no.~5, pp.
  2390--2412, 11 2009. [Online]. Available:
  \url{https://www.researchgate.net/publication/38072230_A_phenomenological_model_of_the_synapse_between_the_inner_hair_cell_and_auditory_nerve_Long-term_adaptation_with_power-law_dynamics}
\BIBentrySTDinterwordspacing

\bibitem{Bruce2003AnResponses}
I.~C. Bruce, M.~B. Sachs, and E.~D. Young, ``{An auditory-periphery model of
  the effects of acoustic trauma on auditory nerve responses},'' \emph{The
  Journal of the Acoustical Society of America}, vol. 113, no.~1, pp. 369--388,
  1 2003.

\bibitem{Wu2019PrimaryEar}
P.~Z. Wu, L.~D. Liberman, K.~Bennett, V.~de~Gruttola, J.~T. O'Malley, and M.~C.
  Liberman, ``{Primary Neural Degeneration in the Human Cochlea: Evidence for
  Hidden Hearing Loss in the Aging Ear},'' \emph{Neuroscience}, vol. 407, pp.
  8--20, 5 2019.

\bibitem{Wu2021PrimaryScores}
P.~Z. Wu, J.~T. O’Malley, V.~de~Gruttola, and M.~C. Liberman, ``{Primary
  neural degeneration in noise-exposed human cochleas: Correlations with outer
  hair cell loss and word-discrimination scores},'' \emph{Journal of
  Neuroscience}, vol.~41, no.~20, pp. 4439--4447, 5 2021.

\bibitem{Drakopoulos2023ACompensation}
F.~Drakopoulos and S.~Verhulst, ``{A Neural-Network Framework for the Design of
  Individualised Hearing-Loss Compensation},'' \emph{IEEE/ACM Transactions on
  Audio Speech and Language Processing}, vol.~31, pp. 2395--2409, 2023.

\bibitem{GitHubHearingTechnology/CoNNear_cochlea}
\BIBentryALTinterwordspacing
``{GitHub - HearingTechnology/CoNNear{\_}cochlea}.'' [Online]. Available:
  \url{https://github.com/HearingTechnology/CoNNear_cochlea}
\BIBentrySTDinterwordspacing

\bibitem{Rahaman2018OnNetworks}
\BIBentryALTinterwordspacing
N.~Rahaman, A.~Baratin, D.~Arpit, F.~Draxlcr, M.~Lin, F.~A. Hamprecht,
  Y.~Bengio, and A.~Courville, ``{On the Spectral Bias of Neural Networks},''
  \emph{36th International Conference on Machine Learning, ICML 2019}, vol.
  2019-June, pp. 9230--9239, 6 2018. [Online]. Available:
  \url{https://arxiv.org/abs/1806.08734v3}
\BIBentrySTDinterwordspacing

\bibitem{Cohen2016GroupNetworks}
\BIBentryALTinterwordspacing
T.~S. Cohen and M.~Welling, ``{Group Equivariant Convolutional Networks},''
  \emph{33rd International Conference on Machine Learning, ICML 2016}, vol.~6,
  pp. 4375--4386, 2 2016. [Online]. Available:
  \url{https://arxiv.org/abs/1602.07576v3}
\BIBentrySTDinterwordspacing

\bibitem{Stoller2018}
D.~Stoller, S.~Ewert, and S.~Dixon, ``{Wave-U-Net: A multi-scale neural network
  for end-to-end audio source separation},'' \emph{Proceedings of the 19th
  International Society for Music Information Retrieval Conference, ISMIR
  2018}, pp. 334--340, 2018.

\bibitem{He2015}
K.~He, X.~Zhang, S.~Ren, and J.~Sun, ``{Delving deep into rectifiers:
  Surpassing human-level performance on imagenet classification},''
  \emph{Proceedings of the IEEE International Conference on Computer Vision},
  vol. 2015 Inter, pp. 1026--1034, 2015.

\bibitem{Zen2019LibriTTS:Text-to-Speech}
\BIBentryALTinterwordspacing
H.~Zen, V.~Dang, R.~Clark, Y.~Zhang, R.~J. Weiss, Y.~Jia, Z.~Chen, and Y.~Wu,
  ``{LibriTTS: A Corpus Derived from LibriSpeech for Text-to-Speech},''
  \emph{Proceedings of the Annual Conference of the International Speech
  Communication Association, INTERSPEECH}, vol. 2019-September, pp. 1526--1530,
  4 2019. [Online]. Available: \url{https://arxiv.org/abs/1904.02882v1}
\BIBentrySTDinterwordspacing

\bibitem{Oppenheim1998DiscreteEdition}
A.~V. Oppenheim and R.~W. Schafer, ``{Discrete Time Signal Processing 2nd
  Edition},'' 1998.

\bibitem{Welch1967}
P.~D. Welch, ``{The Use of Fast Fourier Transform for the Estimation of Power
  Spectra},'' \emph{IEEE Transactions on audio and electroacoustics}, no.~2,
  pp. 70--73, 1967.

\bibitem{Drakopoulos2023AAll}
F.~Drakopoulos, A.~Van Den~Broucke, and S.~Verhulst, ``{A DNN-Based Hearing-Aid
  Strategy For Real-Time Processing: One Size Fits All},'' \emph{ICASSP, IEEE
  International Conference on Acoustics, Speech and Signal Processing -
  Proceedings}, pp. 1--5, 5 2023.

\bibitem{Drakopoulos2023}
\BIBentryALTinterwordspacing
F.~Drakopoulos and S.~Verhulst, ``{A Neural-Network Framework for the Design of
  Individualised Hearing-Loss Compensation},'' \emph{IEEE/ACM Transactions on
  Audio Speech and Language Processing}, vol.~31, pp. 2395--2409, 7 2023.
  [Online]. Available: \url{http://arxiv.org/abs/2207.07091}
\BIBentrySTDinterwordspacing

\bibitem{Bishop2006}
C.~M. Bishop, \emph{{Pattern Recoginiton and Machine Learning}}.\hskip 1em plus
  0.5em minus 0.4em\relax Springer-Verlag New York, 2006.

\bibitem{Bisgaard2010}
\BIBentryALTinterwordspacing
N.~Bisgaard, M.~S. Vlaming, and M.~Dahlquist, ``{Standard Audiograms for the
  IEC 60118-15 Measurement Procedure},'' \emph{Trends in Amplification},
  vol.~14, no.~2, pp. 113--120, 2010. [Online]. Available:
  \url{http://tia.sagepub.com}
\BIBentrySTDinterwordspacing

\bibitem{Luo2018Conv-TasNet:Separation}
\BIBentryALTinterwordspacing
Y.~Luo and N.~Mesgarani, ``{Conv-TasNet: Surpassing Ideal Time-Frequency
  Magnitude Masking for Speech Separation},'' \emph{IEEE/ACM Transactions on
  Audio Speech and Language Processing}, vol.~27, no.~8, pp. 1256--1266, 9
  2018. [Online]. Available: \url{http://arxiv.org/abs/1809.07454
  http://dx.doi.org/10.1109/TASLP.2019.2915167}
\BIBentrySTDinterwordspacing

\bibitem{Tzinis2020SudoSeparation}
\BIBentryALTinterwordspacing
E.~Tzinis, Z.~Wang, and P.~Smaragdis, ``{Sudo rm -rf: Efficient Networks for
  Universal Audio Source Separation},'' \emph{IEEE International Workshop on
  Machine Learning for Signal Processing, MLSP}, vol. 2020-September, 7 2020.
  [Online]. Available: \url{http://arxiv.org/abs/2007.06833
  http://dx.doi.org/10.1109/MLSP49062.2020.9231900}
\BIBentrySTDinterwordspacing

\bibitem{Roux2018SDRDone}
\BIBentryALTinterwordspacing
J.~L. Roux, S.~Wisdom, H.~Erdogan, and J.~R. Hershey, ``{SDR - half-baked or
  well done?}'' \emph{ICASSP, IEEE International Conference on Acoustics,
  Speech and Signal Processing - Proceedings}, vol. 2019-May, pp. 626--630, 11
  2018. [Online]. Available: \url{https://arxiv.org/abs/1811.02508v1}
\BIBentrySTDinterwordspacing

\bibitem{Carney2015}
\BIBentryALTinterwordspacing
L.~H. Carney, T.~Li, and J.~M. McDonough, ``{Speech Coding in the Brain:
  Representation of Vowel Formants by Midbrain Neurons Tuned to Sound
  Fluctuations},'' \emph{eneuro}, vol.~2, no.~4, pp. 0004--15, 7 2015.
  [Online]. Available:
  \url{http://eneuro.org/lookup/doi/10.1523/ENEURO.0004-15.2015}
\BIBentrySTDinterwordspacing

\bibitem{Lybarger1944}
S.~Lybarger, ``{U.S. Patent application SN 543, 278. July 3.}'' 1944.

\bibitem{Nielsen2011TheTest}
\BIBentryALTinterwordspacing
J.~B. Nielsen and T.~Dau, ``{The Danish hearing in noise test},''
  \emph{International Journal of Audiology}, vol.~50, no.~3, pp. 202--208, 3
  2011. [Online]. Available:
  \url{https://www.tandfonline.com/doi/abs/10.3109/14992027.2010.524254}
\BIBentrySTDinterwordspacing

\bibitem{RadiocommunicationSector2015Methodsound}
\BIBentryALTinterwordspacing
I.~Radiocommunication~Sector, ``{Method for the subjective assessment of
  intermediate quality level of audio systems BS Series Broadcasting service
  (sound)},'' 2015. [Online]. Available:
  \url{http://www.itu.int/ITU-R/go/patents/en}
\BIBentrySTDinterwordspacing

\bibitem{Keidser2011}
G.~Keidser, H.~Dillon, M.~Flax, T.~Ching, and S.~Brewer, ``{The NAL-NL2
  Prescription Procedure},'' \emph{Audiology Research}, vol.~1, no.~1, p. e24,
  2011.

\bibitem{Nielsen2009}
J.~B. Nielsen, \emph{{Assessment of speech intelligibility in background noise
  and reverberation}}.\hskip 1em plus 0.5em minus 0.4em\relax Technical
  University of Denmark, 2009.

\bibitem{Raudenbush2001}
A.~S. Bryk and S.~W. Raudenbush, \emph{{Hierarchical linear models:
  Applications and data analysis methods. Advanced quantitative techniques in
  the social sciences}}.\hskip 1em plus 0.5em minus 0.4em\relax SAGE
  Publications, 1992.

\bibitem{Studebaker1985ATransform}
\BIBentryALTinterwordspacing
G.~A. Studebaker, ``{A "Rationalized" Arcsine Transform},'' \emph{Journal of
  speech and hearing research}, vol.~28, no.~3, pp. 455--462, 1985. [Online].
  Available:
  \url{https://pubs.asha.org/doi/10.1044/jshr.2803.455?url_ver=Z39.88-2003&rfr_id=ori%3Arid%3Acrossref.org&rfr_dat=cr_pub++0pubmed}
\BIBentrySTDinterwordspacing

\bibitem{SHAPIRO1965AnSamples}
\BIBentryALTinterwordspacing
S.~S. SHAPIRO and M.~B. WILK, ``{An analysis of variance test for normality
  (complete samples)},'' \emph{Biometrika}, vol.~52, no. 3-4, pp. 591--611, 12
  1965. [Online]. Available: \url{https://dx.doi.org/10.1093/biomet/52.3-4.591}
\BIBentrySTDinterwordspacing

\bibitem{Anderson1954AFit}
T.~W. Anderson and D.~A. Darling, ``{A Test of Goodness of Fit},''
  \emph{Journal of the American Statistical Association}, vol.~49, no. 268, pp.
  765--769, 1954.

\bibitem{Jarque1987AResiduals}
C.~M. Jarque and A.~K. Bera, ``{A Test for Normality of Observations and
  Regression Residuals},'' \emph{International Statistical Review / Revue
  Internationale de Statistique}, vol.~55, no.~2, p. 163, 8 1987.

\bibitem{Lopez-Espejo2023OnIntelligibility}
I.~L{\'{o}}pez-Espejo, A.~Edraki, W.~Y. Chan, Z.~H. Tan, and J.~Jensen, ``{On
  the deficiency of intelligibility metrics as proxies for subjective
  intelligibility},'' \emph{Speech Communication}, vol. 150, pp. 9--22, 5 2023.

\bibitem{Gonzalez2023AssessingEnvironments}
P.~Gonzalez, T.~S. Alstr{\o}m, and T.~May, ``{Assessing the Generalization Gap
  of Learning-Based Speech Enhancement Systems in Noisy and Reverberant
  Environments},'' \emph{IEEE/ACM Transactions on Audio, Speech, and Language
  Processing}, 2023.

\end{thebibliography}

\end{document}